# Design of a medium energy, high average power superconducting e-beam accelerator for environmental applications

R.C. Dhuley[1], I. Gonin[1], K. Zeller[2], R. Kostin[3], S. Kazakov[1], B. Coriton[2], T. Khabiboulline[1], A. Sukhanov[1], V. Yakovlev[1], A. Saini[1], N. Solyak[1], A. Sauers[1], J.C.T. Thangaraj[1]

[1]Fermi National Accelerator Laboratory, Batavia IL, 60510, United States

[2]General Atomics, San Diego CA, 92186, United States

[3]Euclid Techlabs, LLC, Bolingbrook IL, 60440, United States

## Abstract

We present the technical and engineering design of a medium energy (10 MeV) and high average power (1 MW) electron beam accelerator intended for irradiation treatment of high volume industrial and municipal wastewater. The accelerator uses superconducting radiofrequency (SRF) cavity technology for producing the high average beam power with >90% RF to beam efficiency. The design of the accelerator is simplified and tailored for industrial settings by adopting the cryocooler conduction-cooling technique for the SRF cavity instead of a conventional liquid helium bath cryo-system. The technical design is supplemented with a detailed analysis of capital and operating cost of the accelerator. The designed accelerator can treat up to 12 million gallons per day of wastewater, requires capital of ~$8M for construction, and has ~13.5 ¢/ton/kGy in material processing cost.

## 1. Introduction

Energetic electron beams (e-beams) are a powerful tool for numerous applications ranging from scientific R&D to industrial processes. E-beams with GeV energy are used to probe matter for particle physics research and generate powerful x-rays for photon sciences research [1]. MeV-scale (1-10s of MeV) e-beams are a material processing tool for altering the physical, chemical, molecular, and biological properties of materials. These include polymerization, medical and food sterilization, environmental remediation, wastewater treatment, sludge and biosolids treatment, etc. [2]. X-rays generated using MeV-scale e-beams find application in cargo scanning and to generate THz light for security imaging applications [3]. More recently, ultra-short, high-quality MeV-scale electron beam are opening opportunities into ultrafast electron microscopy/diffraction applications [4]. With more than 1400 industrial installations, electron beam accelerators remain as the primary source of energetic electron beams for the above industrial processes [5].

Following its successful demonstration on several pilot projects [6], electron irradiation has garnered recent attention for high-volume applications such as municipal and industrial wastewater and sludge treatment. E-beam based water treatment was successfully demonstrated as early as 1988 at the Miami-Dade Virginia Key Wastewater Treatment Plant, where the process utilizing a 1.5 MeV e-beam at 50 mA (75 kW average power) successfully disinfected anaerobically digested sludge [7]. The plant achieved >99% removal efficiency for some organic compounds and ~77% removal efficiencies for most compounds. In 1997, a pilot-scale e-beam treatment facility was commissioned at Daegu Dyeing Industrial Complex (DDIC) in South Korea. Initially specified for 1 MeV and 40 kW, the power was upgraded in 2005 to 400 kW. At 400 kW, the plant removed dye from 10,000 $m^3$ of wastewater per day [8]. In 2015, increased environmental regulation in Jiangsu Province, China led a wastewater treatment plant to investigate more advanced effluent remediation techniques. Proof-of-principle studies were performed with a Rhodotron TT200 electron accelerator at 10 MeV and 10 mA (100 kW of average power) [9]. Following success of this pilot, China opened the world's largest wastewater treatment facility at Guanhua Knitting Factory in Southern

China. Though details on the e-beam energy, current, and power have not yet been publicized as of the date of this paper, the facility is purportedly able to treat 30 million liters of industrial wastewater per day and save 4.5 billion liters of fresh water annually [10].

A recent study conducted at Fermilab [11] determined the requirements for 10 MeV e-beam power to treat 2 million gallons (~8 million liters) of wastewater per day at the Metropolitan Water Reclamation District (MWRD) of Greater Chicago, one of the largest municipal wastewater treatment facilities in the United States. Nearly 1 MW of e-beam was deemed sufficient for treating dewatered biosolid sludge or the pre-anaerobic digester thickened Waste Activated Sludge (WAS) stream in the MWRD Stickney Plant at 2 Million Gallons per Day (MGD). For the higher flow-rate of 8-13 MGD of wastewater encountered upstream of WAS Thickener where there is a great opportunity to treat and recover water, e-beam power more than 5 MW is required. Both these applications can be well-served by accelerator units delivering MW-scale of 10 MeV e-beam.

Linear accelerators (linacs) using the room temperature copper RF cavities are attractive for the MeV-scale energy range required for wastewater and sludge treatment. However, these normal conducting RF cavities are constrained to operate at very low RF duty cycle due to high RF heating at the cavity walls. This low RF duty operation usually limits the average e-beam power obtainable from normal conducting linacs to few 10s of kWs, making them less viable for high-volume irradiation applications. Superconducting RF cavities made of pure niobium or $Nb_3Sn$, with cryogenic operation near the temperature of 4 K, exhibit extremely small RF wall dissipation (about six orders of magnitude smaller than copper cavities of comparable shape and size), allowing their operation at 100% RF duty cycle (continuous wave or cw operation). SRF cavities can thus produce very high average power e-beams suitable for high-volume irradiation applications. It is also well known that SRF cavities can be more energy and cost efficient compared to copper cavities even after accounting for the energy and cost premium required for their cryogenic operation [12].

A prior publication by Ciovati *et al.* [13] reports a design of a 1 MeV, 1 MW SRF based e-beam accelerator for the treatment of flue gases and wastewater. While the energy level of 1 MeV selected by Ciovati *et al.* is more suited for flue gas - a low density material, a 10 MeV e-beam is more practical for treating higher density materials such as wastewater and sludge. This is because the penetration depth of 10 MeV electrons in water is 10-fold of 1 MeV electrons, which enables treating larger volume flows of water per unit time. Motivated by this striking advantage, we have designed a 10 MeV, 1 MW e-beam accelerator for high-volume (>MGD) wastewater treatment. The design and economic assessment (capital and operating expense) of this accelerator is the prime subject of the present paper.

The accelerator has a pre-accelerator powered by a room temperature electron source and an injector cavity. The main accelerator uses a scalable cryogenic module (cryomodule) in which a SRF cavity is conduction-cooled using closed-cycle 4 K cryocoolers. Unlike conventional SRF cryo-systems (example [14]), this technique makes use of neither large-scale helium cryogenic infrastructure nor complex liquid helium containing cryomodules. The technique offers the advantages of operational safety (less stringent loss of beamline vacuum [15,16]), simpler construction (simpler pressure vessel and pressure relief system), and reliability that are attractive for industrial settings.

This paper is structured to start with the design of the pre-injector followed by the SRF cryomodule, including detailed beam dynamics simulations for attaining the 10 MeV, 1 MW final beam. We keep focus on the component-level engineering design of the SRF cryomodule, its assembly procedure, and then capital cost estimation. Finally, the accelerator wall plug efficiency and operating expense are estimated to evaluate the expected cost of wastewater treatment using beams produced by this accelerator.

## 2. Accelerator design

### a. Accelerator layout and components

Figure 1 depicts the major accelerator components and their layout. The layout is divided into three sections: pre-accelerator, accelerator (also referred to as cryomodule), and beam delivery system. The pre-accelerator is comprised of thermionic electron source (gun), an RF injector cavity, and a focusing solenoid magnet. The electron beam exiting the pre-accelerator is fed into the accelerator, which energizes the beam to the 10 MeV target energy. The accelerator uses a $Nb_3Sn$ (SRF) cavity operating near 4 K, conduction cooled by a bank of cryocoolers. The cavity cold mass is enclosed in a 50 K thermal intercept shield, surrounded by a room temperature magnetic shield. The cavity cold mass and the two shields are housed in a vacuum vessel at room temperature. Two fundamental power couplers pierce the vacuum vessel through two ports at 180 degrees to each other, to feed RF power into the SRF cavity. The beam exits the accelerator with 10 MeV energy and then enters the beam delivery system where it is conditioned using a raster magnet and beam horn for irradiating a stream of wastewater. The electron beam accelerator is ~4 m long (end-to-end), ~2 m width, and ~2 m tall.

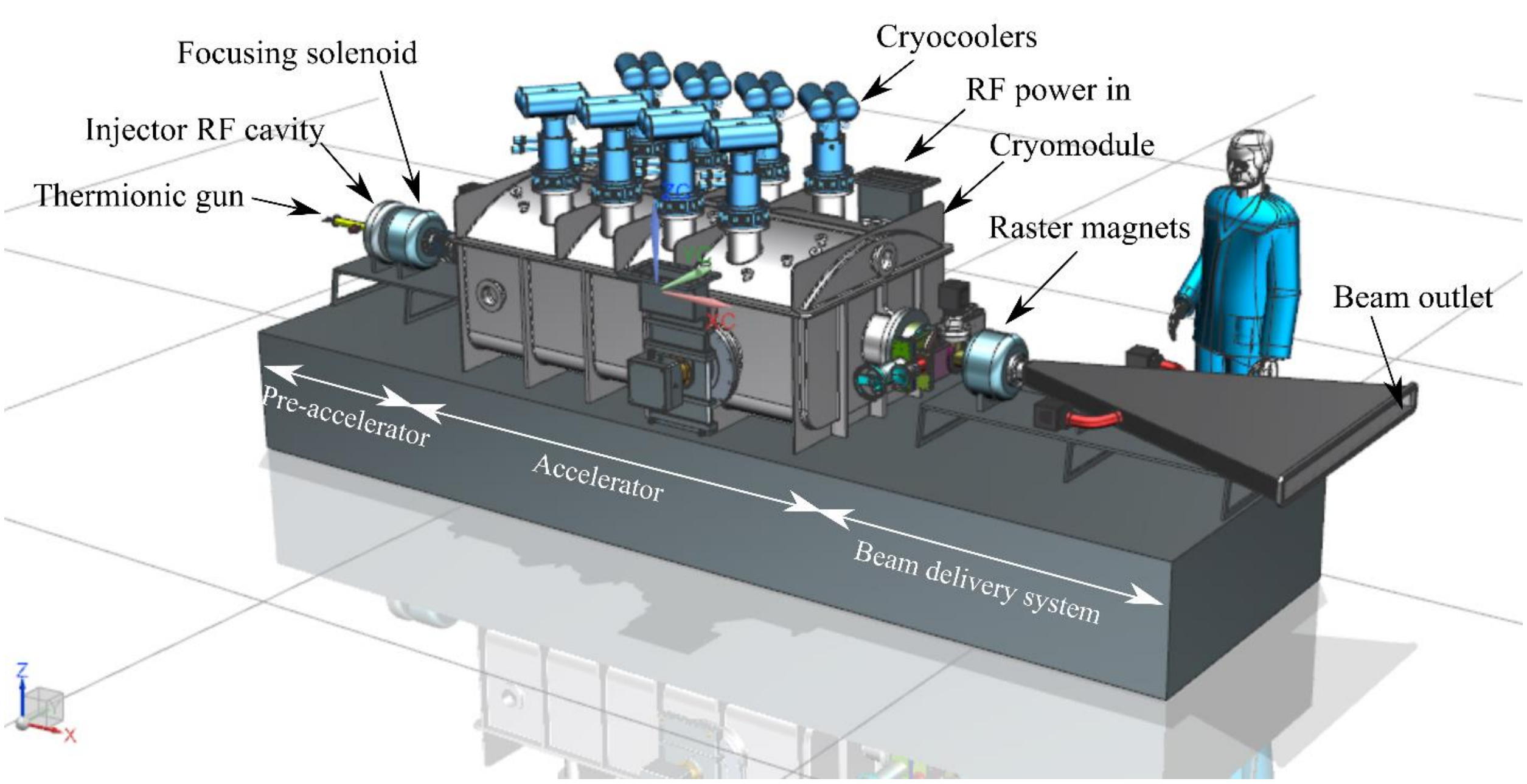

*Figure 1: E-beam accelerator components and layout. The overall size is ~4 m long (end-to-end), ~2 m wide, and ~2 m tall.*

### b. Pre-accelerator

The pre-accelerator is composed of an electron gun, an injector cavity, and a focusing solenoid magnet arranged in the stated order. These components are designed to operate at room temperature and are situated outside of the cryomodule.

#### i. *Electron source (gun)*

The pre-accelerator herein uses a triode RF gun with a gridded thermionic cathode. In this gun the cathode emits low energy electrons *via* thermal emission, which are then shaped into electron bunches using the RF voltage applied to the grid-cathode gap, superimposed on a constant DC voltage. The emitted electrons are then captured and accelerated by electric field of the RF gun. The operating RF amplitude and phase interval for the gun are determined for producing 100 mA average current with 154 pC electron beam bunch charge. 3D particle tracker software MICHELLE is used for optimizing beam emittance, energy spread, and RMS bunch length at the gun exit. The parameter summary of cathode and beam after the grid are presented in Figure 2.

Also shown in Figure 2 is a cross-section of the RF gun and its main components. Internal structural of the RF Gun has three detachable parts: gun RF resonator with power coupler, thermionic cathode, and grid assembly. In the operating position the grid's outside surface is directly facing the accelerating gap entrance of the injector cavity (described in the following section). The cathode unit is mounted to the RF gun resonator by a flanged connection and can be separated from the gun for maintenance. The standard series barium tungsten dispenser cathode with diameter of 12.7 mm and operating temperature of 950-1200 ºC is considered for the present study. A bellows is used as part of the outer conductor of the RF gun for mechanical adjustment of the cathode-grid distance.

| Parameters | Unit | Value |
|---|---|---|
| **Frequency** | MHz | 650 |
| **Cathode diameter** | inch | 0.5 |
| **Beam current** | mA | 100 |
| **Current density** | A/cm$^2$ | 2.35 |
| **DC bias voltage** | kV | 2.6 |
| **Output Energy** | keV | 3.5 |
| **Bunch rms size** | Deg | <15 |
| **Energy rms size** | % | <25 |

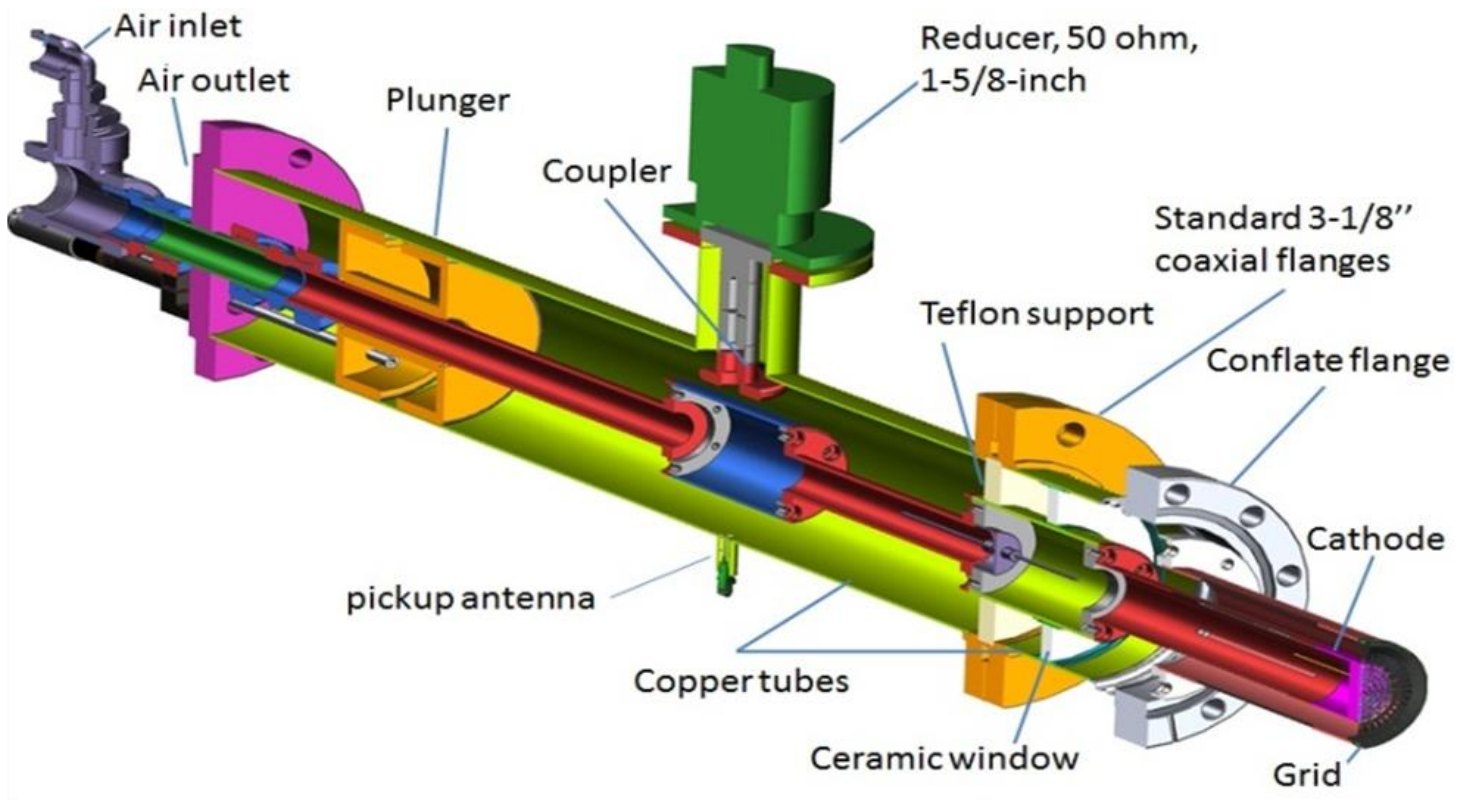

*Figure 2: RF gun design parameters (left) and components (right).*

*ii. Injector cavity RF design*

The injector cavity, located immediately downstream of the RF gun, captures the thermionically emitted electrons, and accelerates them to ~300 keV energy. RF design of the injector cavity is done using CST Microwave Studio software. The goal of cavity RF design is to maximize the shunt impedance to get the required accelerating voltage with minimum heat dissipation. The main dimensions for optimization are longitudinal length, $L_{cavity}$, accelerating gap length, $L_{gap}$, and the radii $R_1$ and $R_2$ as depicted in Figure 3. The cavity diameter is chosen to attain TM010 mode resonance at 650 MHz. The optimized geometrical dimensions and the resulting RF parameters are also summarized in Figure 3.

Taking copper as the injector cavity material, the voltage gain of ~300 kV would dissipate 11.6 kW of heat, which will be extracted using forced flow of cooling water around the cavity.

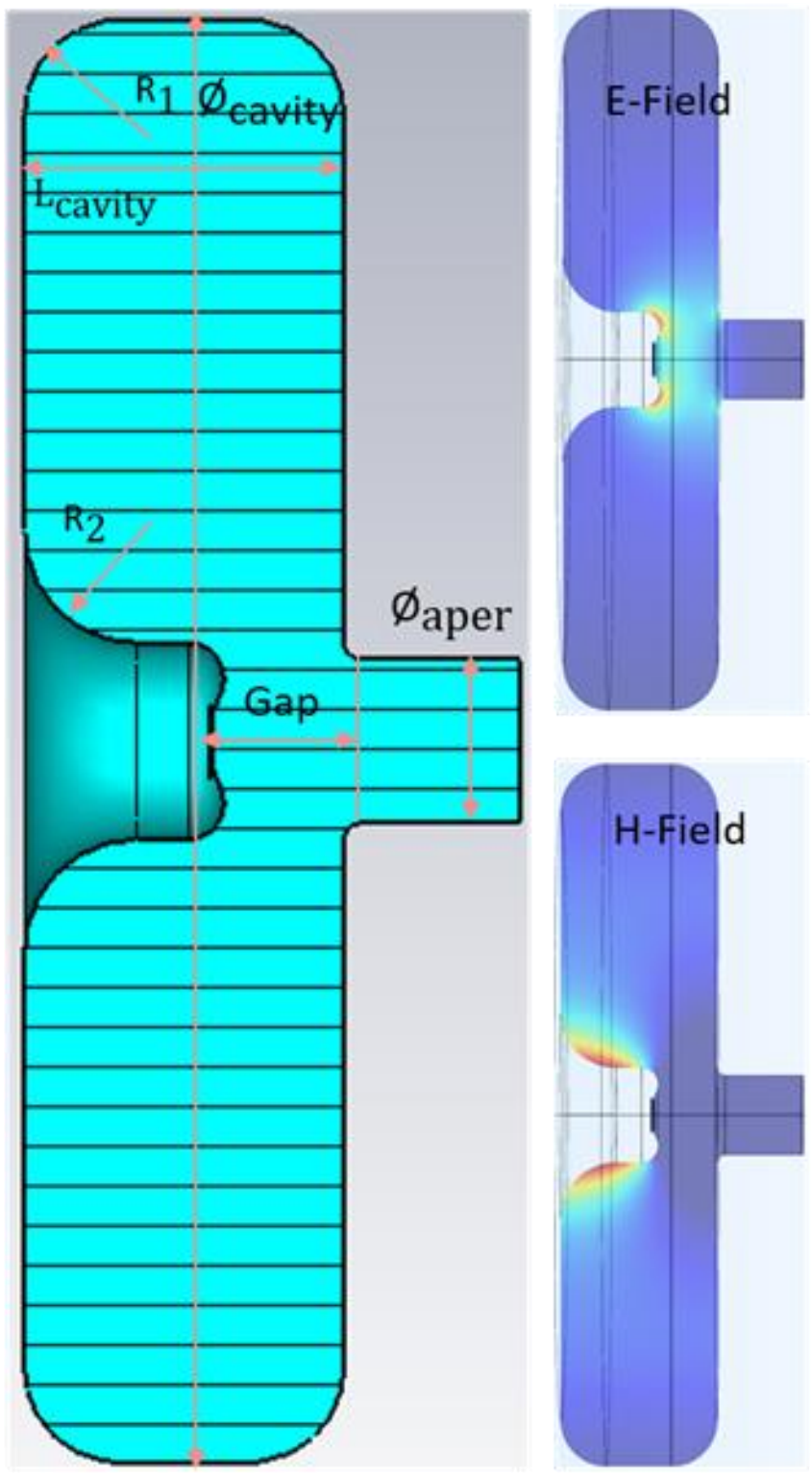

| Parameter | Value, mm |
|---|---|
| Frequency | 650 MHz |
| Øcavity | 308 |
| Øaperture | 35 |
| $R_1$ | 20 |
| $R_2$ | 24 |
| Gap | 29.6 |
| $L_{cavity}$ | 68.2 |

| Parameter | Value |
|---|---|
| V, kV | 200 |
| R/Q, Ω | 178 |
| Q | 19000 |
| $R_{sh}$, MΩ | 3.4 |
| Losses, kW | 11.6 |
| $E_s$, max, MV/m | 14.5 |

*Figure 3: Injector cavity design and operating parameters.*

### *iii. Beam dynamics simulations of the RF gun and injector cavity*

The simulation of the electron emission for the cathode-grid region with the RF gun resonator was carried out with MICHELLE. The control cathode voltage has the following time dependence:

$$U(t) = U_d + U_a \cos(\omega t + \phi) \quad (1)$$

where $U_d$ is constant bias voltage, $U_a$ is amplitude of the bias RF voltage, $\omega$ is RF frequency and $\phi$ is phase shift between bias RF field in the injector and RF field in the gun cavity. For the optimization we varied $U_d$, $U_a$, $\phi$ as well as injector voltage to obtain the required average current $I_{avg}$ = 0.1 A and 0.3 MeV beam output energy and to minimize bunch length RMS and beam energy spread RMS. Figure 4 shows MICHELLE plots of emitted particles at the time of the beginning of emission from the cathode surface (left plot) and the β·γ distribution of the beam at the moment of its propagation through the middle of the injector gap (center plot). Optimized parameters are also provided in the table within Figure 4. The simulations use quarter model of the gun owing to symmetry.

The calculated beam characteristics and particle distributions at the exit of the injector cavity are summarized in Figure 5. A beam spot size of ~12 mm diameter is obtained at the exit of the injector cavity.

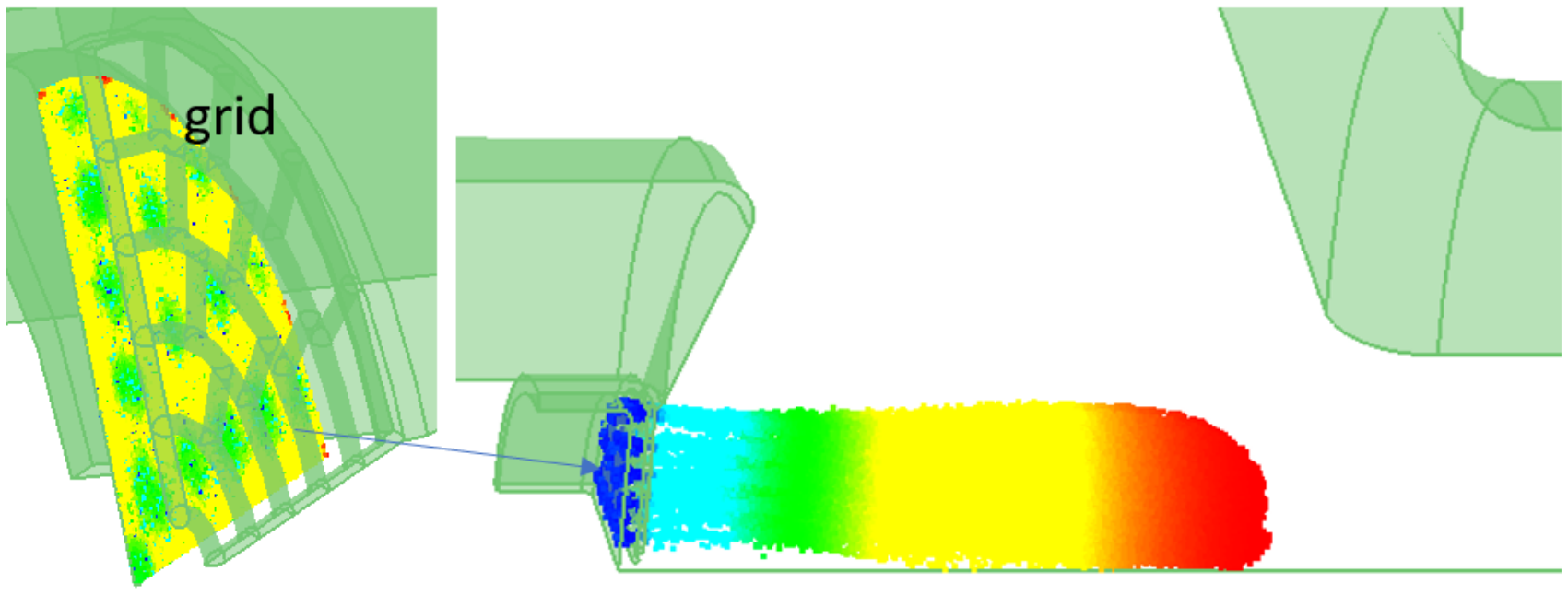

| Parameters | Value | Unit |
|---|---|---|
| Energy | 0.3 | MeV |
| Current | 100 | mA |
| $V_{dc}$ | 2.6 | kV |
| phase shift | 90 | º |
| σ Energy | 7.5 | % |
| σ phase | 7.7 | º |

*Figure 4: Beam distribution at the time of emission from the cathode (left), β·γ beam distribution in the middle of the injector gap (center), table with parameters (right).*

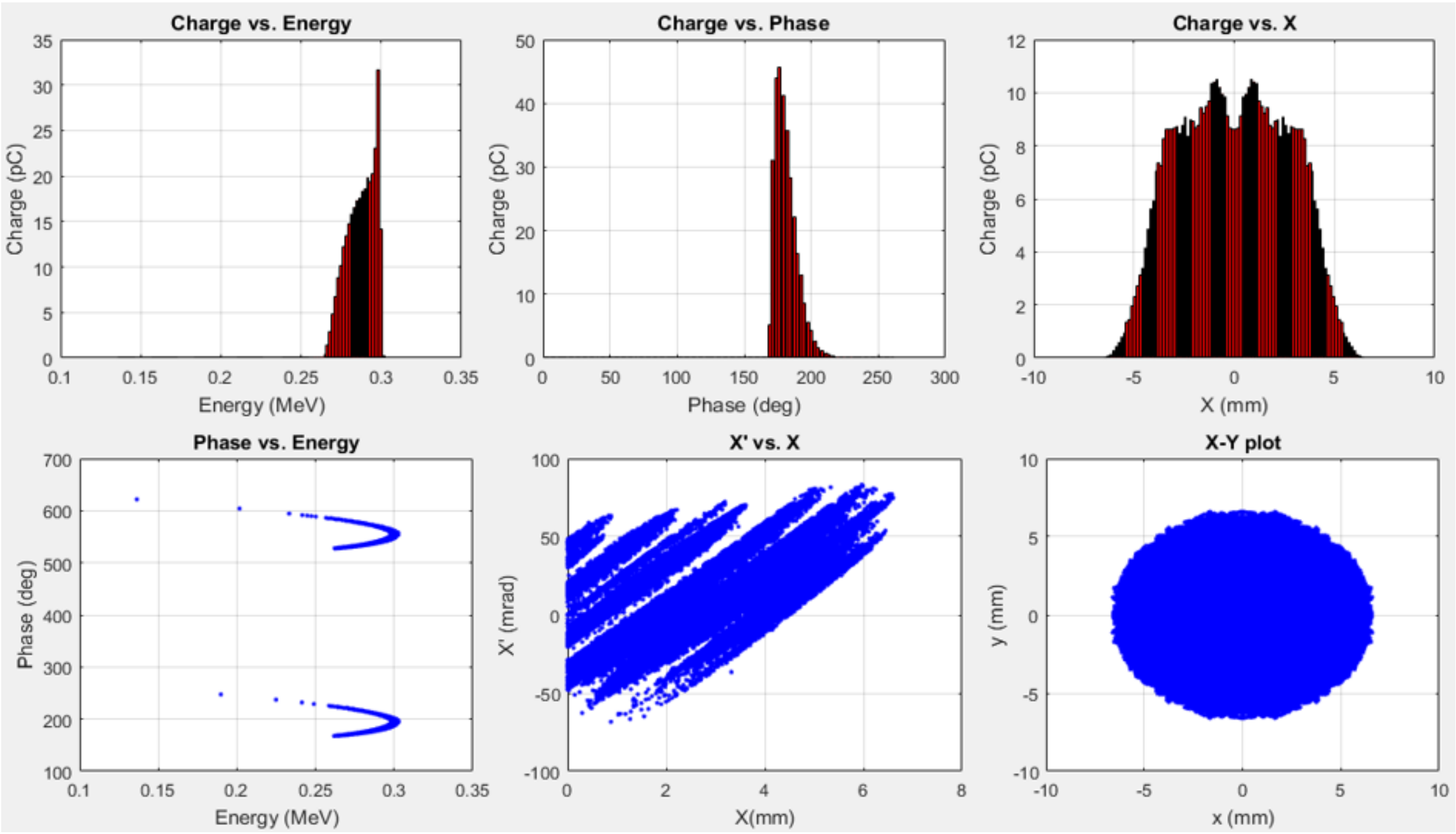

*Figure 5: Beam characteristics at the exit of the injector cavity. Top three plots: charge distribution vs. energy and phase, current vs. radius; Bottom three plots: phase vs. energy, x´-x phase-space distribution, x-y plot at the injector exit.*

### *iv. Focusing solenoid*

The focusing solenoid operating at room temperature is used to match the beam transverse optics to the SRF cavity. The solenoid should (a) provide required focusing properties, (b) be compact, and (c) being place close to the SRF cavity, should create only small remnant magnetic field (a few mG) on the cavity surface to avoid reduction in $Q_0$. The solenoid is placed at the distance of ~300 mm upstream of the SRF cavity to avoid the bunch lengthening, requiring it to provide 0.3 m of focusing distance. This translates to the solenoid having the length of ~100 mm, providing the field of 200-250 G. Hence, focusing strength is of about $6x10^{-5}$ $T^2m$. Table 1 shows the solenoid design parameters that produce the required field profile and focusing strength. The required solenoidal field profile generated by an electromagnetic solenoid with the dimensions in Table 1 is shown in Figure 6.

*Table 1: Parameters for the focusing solenoid at the accelerator inlet.*

| **Solenoid parameter** | **Value** |
|---|---|
| Coil ID/OD | 50/120 mm |
| Coil length | 90 mm |
| Peak field on axis | 0.025 T |
| Current density | 0.4 A/mm$^2$ |
| Focusing strength, Int($B^2$dl) | 6x10$^{-5}$ T$^2$m |

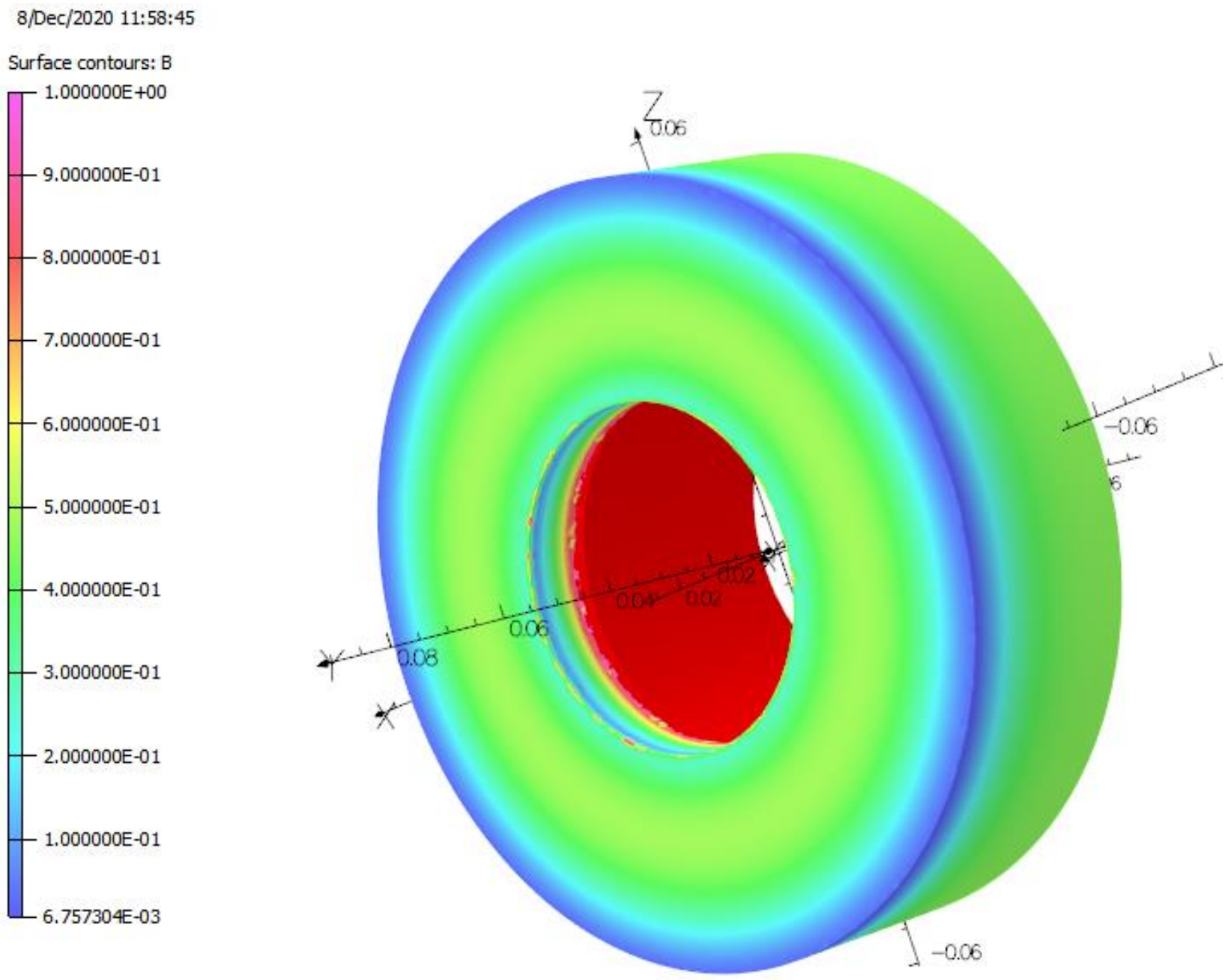

*Figure 6: Surface magnetic field produced by the solenoid magnet.*

## c. Main accelerator cavity RF design and beam transport simulations

### *i. Cavity RF design*

Figure 7 shows geometrical dimensions of the 5-cell, 650 MHz cavity designed to produce the 10 MeV electron beam. The cavity inlet port has 35 mm diameter, equal to that of the injector cavity outlet. This is much larger than the beam spot size of ~12 mm at that location. To match the phase of low-beta electrons entering the cavity, the first cell of the cavity has shorter length compared to the other four cells. The cells 2-4 have the same length and diameter, while the fifth cell is longer and larger in diameter. The outlet iris of the fifth cell and the downstream beampipe are also larger in diameter compared to the other four irises. This larger size is chosen to achieve adequate coupling of the fundamental power coupler with the 5-cell cavity as well as out-propagation of any higher order modes. The outlet beampipe has two coupler ports, placed 180 degrees from each outer, for feeding RF power to the cavity. The beampipe diameter downstream of the coupler port location is reduced to match the diameter of the beam delivery system.

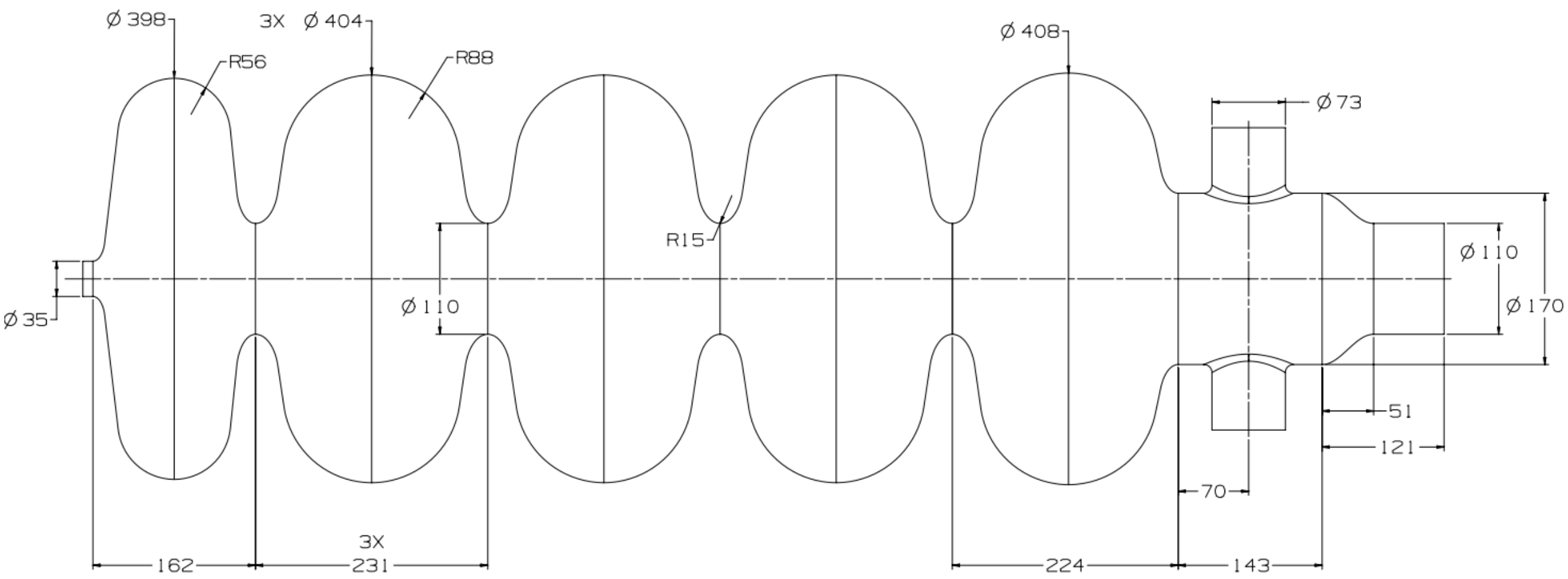

*Figure 7: Geometrical dimensions of the 5-cell cavity.*

The geometrical dimensions shown in Figure 7 are obtained *via* coupled RF and beam transport optimization of the cavity to maximize the quantity *G*R/Q* for the required 10 MeV voltage gain. This maximization ensures minimum heat dissipation for a given surface RF resistance, $R_s$. The optimization also considers obtaining reasonable peak field ratios and good flatness of the surface magnetic field. The RF fields in the TM010 eigenmode are depicted in Figure 8 along with axial electric field and surface magnetic field profiles. The optimization produced uniform axial field profile and flat surface magnetic field profile in each of the five cells. Table 2 lists the optimized RF parameters of the cavity. The total cavity length is 1.35 m of which 1.08 m is the accelerating length.

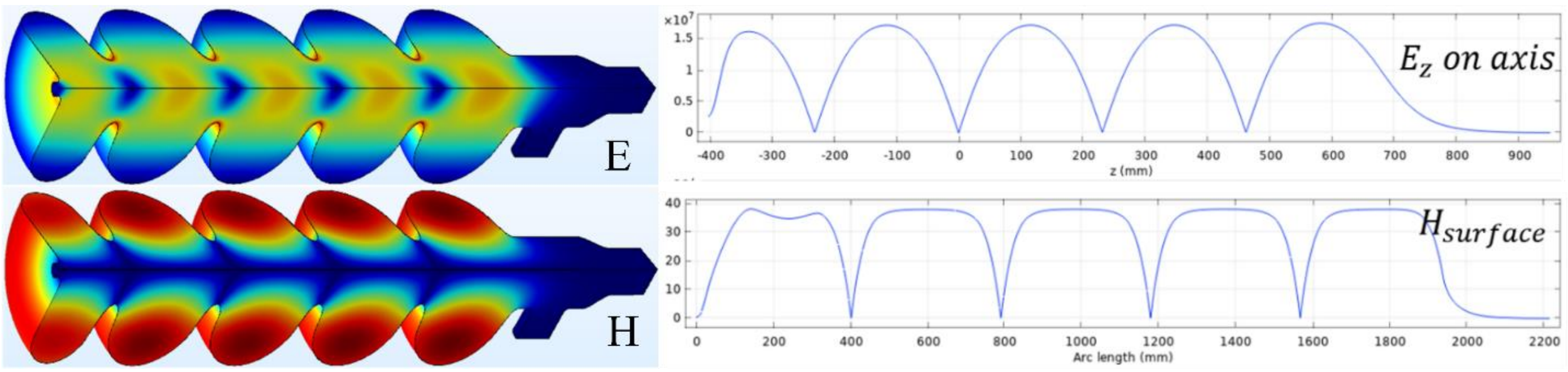

*Figure 8: Electric and magnetic fields distribution on the surface and along the axis.*

*Table 2: Parameters of the optimized 650 MHz 5-cell SRF cavity.*

| Parameter | Value |
|---|---|
| Normalized shunt impedance, $R/Q$ | 635 Ω |
| Geometry factor, $G$ | 262 Ω |
| Dissipated power, $P_{diss}$ at 10 MeV | 0.6* $Rs$ [nΩ] W |
| Peak surface electric field, $E_{s,peak}$ | 17.5 MV/m |
| Peak surface magnetic field, $B_{s,peak}$ | 36.5 mT |

*ii. Coupler side cavity end-group design*

As previously stated, the diameter of the fifth cell and outlet beampipe is enlarged compared to the regular cavity iris from 110 to 170 mm to allow (a) for out-propagation of HOMs and (b) achieve high cavity coupling with the fundamental power couplers. It is also necessary to ensure a sufficiently small operating value of the $Q_{ext}$ ~ 1.5x10$^5$ and simultaneously avoid proximity of the antenna tip to the beam axis. The shape of antenna tip that balances these two opposing requirements is shown in the plot of Figure 9. A 180-degree rotation of the antenna tip brings the $Q_{ext}$ in range of 1-2.5x10$^5$ that includes the target $Q_{ext}$ ~1.5x10$^5$. The antenna tip is oriented 60 deg with the cavity beam axis as depicted in Figure 9 (right) and located 71.5 mm from the beam axis to attain the required $Q_{ext}$. Figure 9 also shows the main dimension of fundamental power coupler location and log scale plot of magnetic field distribution in the cavity end-group.

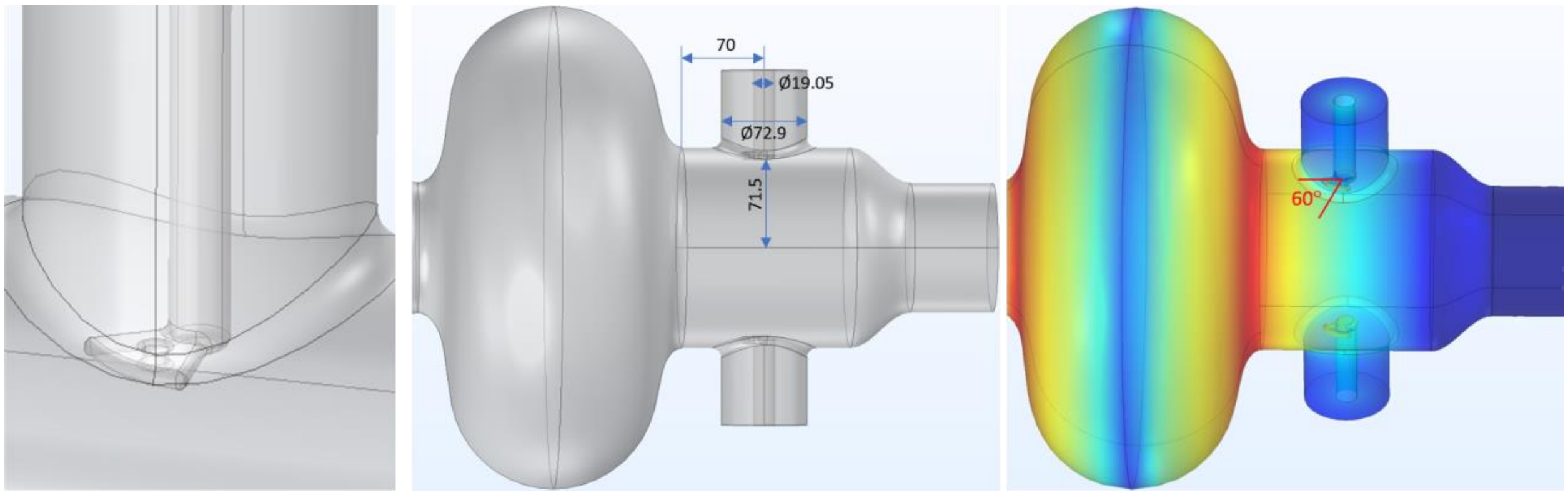

*Figure 9: Zoomed in view of coupler antenna tip (left), dimensions of power coupler port on the cavity and antenna position (center), log scale of the magnetic field distribution in the cavity end-group (right).*

*iii. Simulation of beam transport through the cavity*

The optimized bunch distribution at the output of the injector cavity is used at the entrance of the 5-cell SRF cavity to simulate the bunch acceleration through the 5-cell cavity. This assumes negligible bunch distortion between the output of the injector cavity and inlet of the 5-cell cavity, facilitated by the focusing solenoid. Note that under this assumption we have excluded the focusing solenoid from beam transport simulations. Three sets of fields as shown in Figure 10 are used in MICHELLE beam transport simulations. The amplitude and phase shift in the 5-cell cavity are matched to obtain the 10 MeV beam at the outlet of the accelerator.

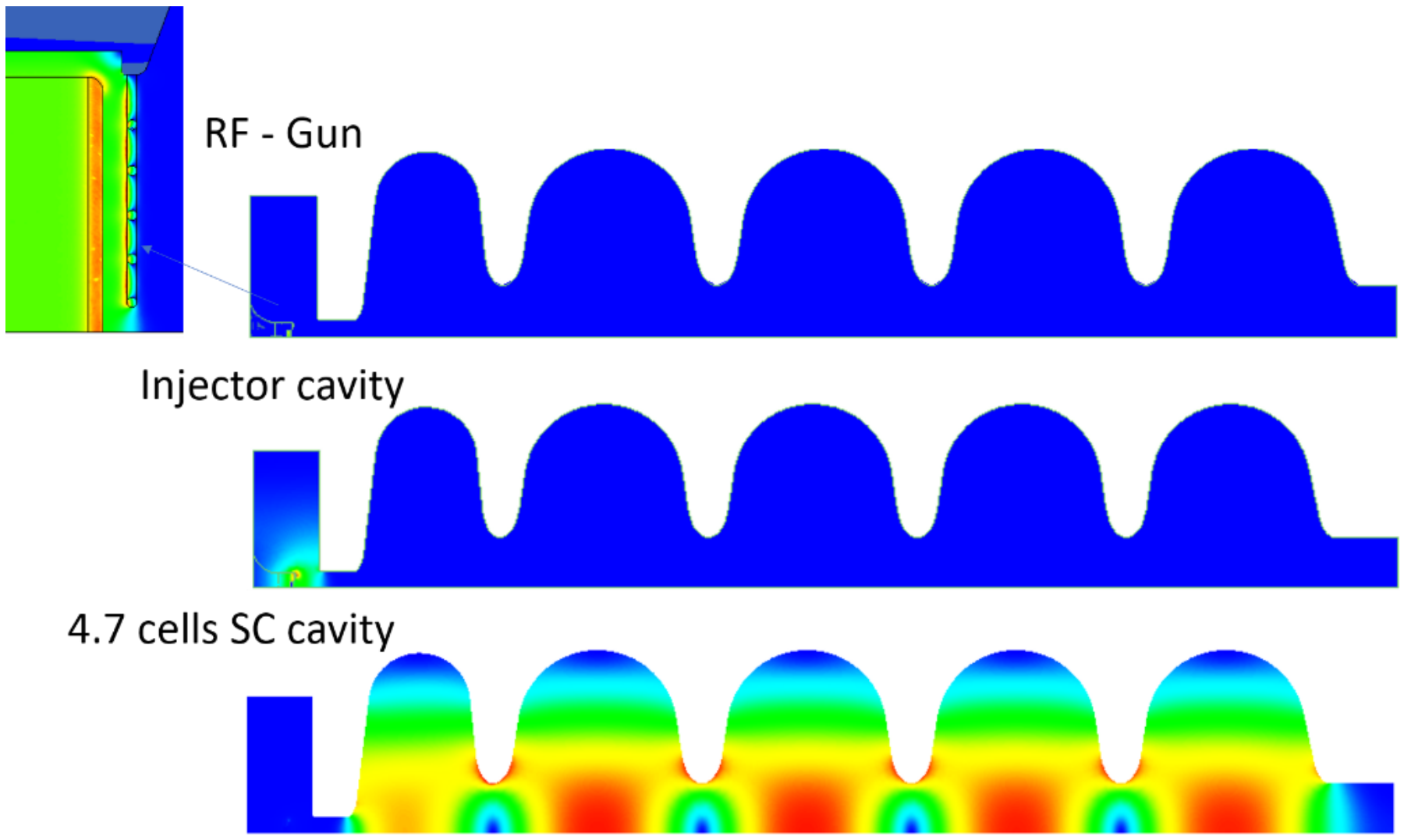

*Figure 10: The three sets of EM-fields used for MICHELLE simulation of beam transport through the accelerator.*

Using up to 100,000 particles and neglecting particle loss to cavity walls, the MICHELLE simulations confirmed that the 5-cell cavity depicted in Figure 7 produces uniform acceleration in each cell, starting from 0.3 MeV at the injector outlet and ending in 10 MeV at the 5-cell cavity outlet. The progressive bunch energy gain through the cavity cells and beam parameters used in the simulation are summarized in Figure 11. Finally, the beam characteristics and particle distributions at the exit of the 5-cell cavity are summarized in the plots of Figure 12.

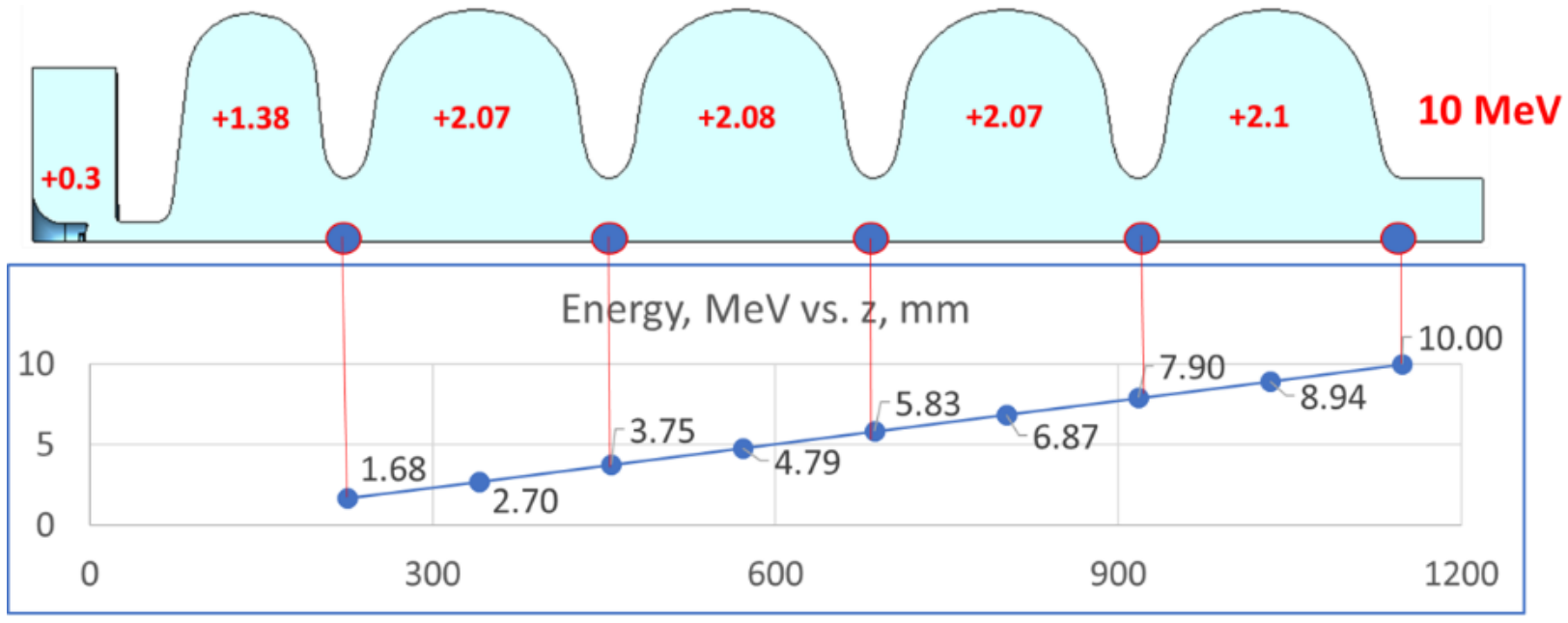

| Frequency | 650 | MHz |
|---|---|---|
| Energy | 10 | MeV |
| Current | 100 | mA |
| phase shift gun-injector | 90 | º |
| phase shift injector-cavity | -15 | º |
| σ Energy | 2.4 | % |
| Beam phase duration rms | 7.5 | º |
| Losses on the cavity walls | 0 | W |

*Figure 11: Beam transport through the 5-cell SRF cavity: cell-by-cell bunch energy gain (left) and output beam parameters (right).*

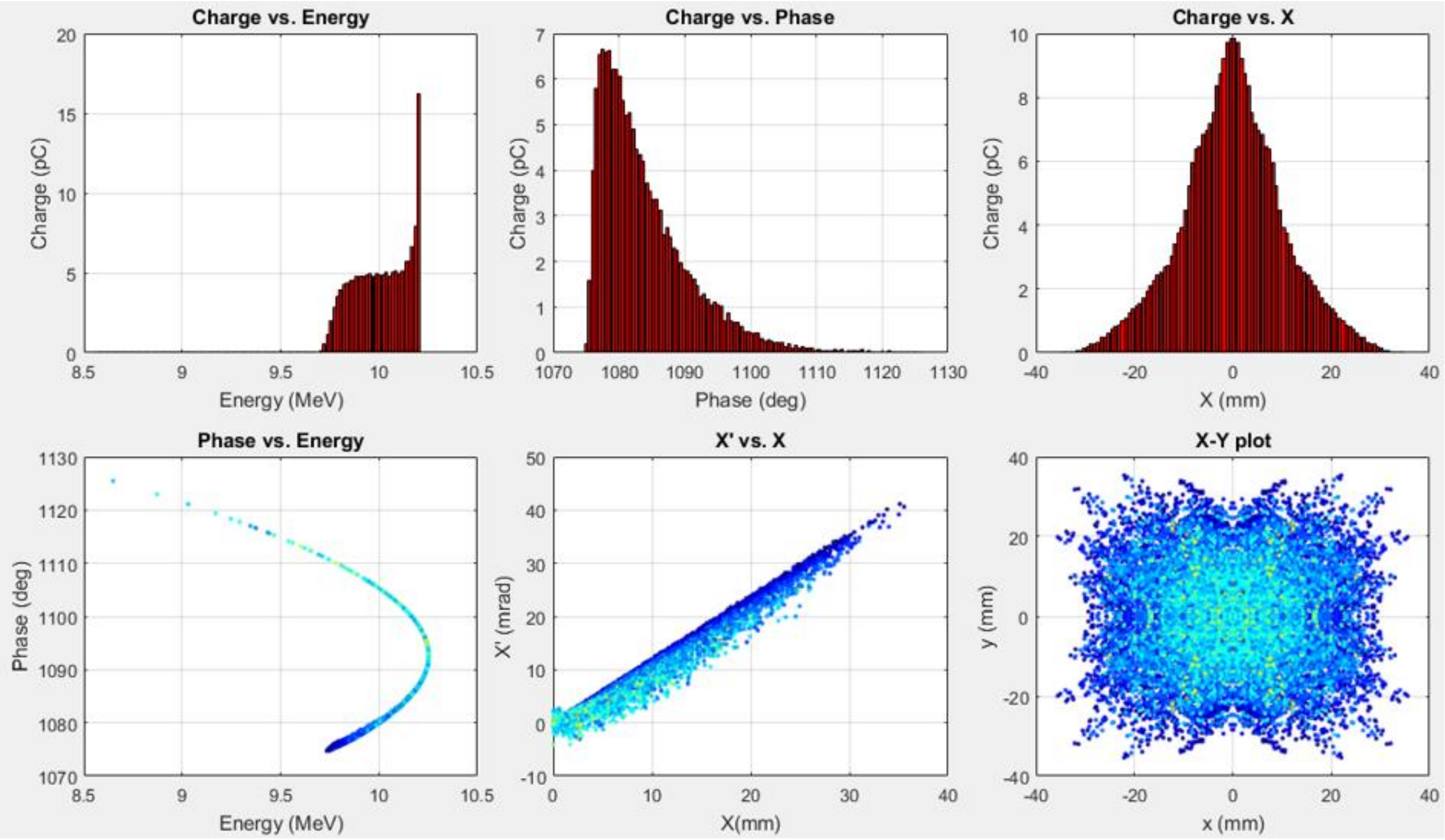

*Figure 12: Beam characteristics and particle distributions at the exit of the 5-cell cavity. Top plots: charge distribution vs. energy and phase, current vs. radius; bottom plots: phase vs. energy, x´-x phase-space distribution, beam spot size.*

### *iv. Analysis of Higher Order Modes (HOMs)*

An understanding and mitigation of cavity higher order modes (HOMs) is essential to ensure proper beam transport through the cavity. In this section, we present the calculated cavity HOM spectrum including monopole and dipole modes and discuss their impact on beam transport through the cavity.

Cavity HOM Spectrum

The monopole and dipole HOM spectrum of the 5-cell cavity shown in Figure 13 are calculated by eigenmode simulations in CST Microwave Studio. Most "dangerous" monopole modes are close to the bunch frequency of 650 MHz. However, these HOMs have only 1% normalized impedance, $(r/Q)_{monopole}$ of that of the fundamental frequency. All other monopole HOMs are far away from harmonic multiples of the fundamental frequency and also have relatively small $(r/Q)_{monopole}$. We therefore conclude that the monopole HOM excitation will not have drastic effect on bunch acceleration. Similarly, all the dipole HOMs are far away from the fundamental of 650 MHz as well as the second harmonic of 1300 MHz. Thus, dipole HOMs excitation is also expected to be insignificant.

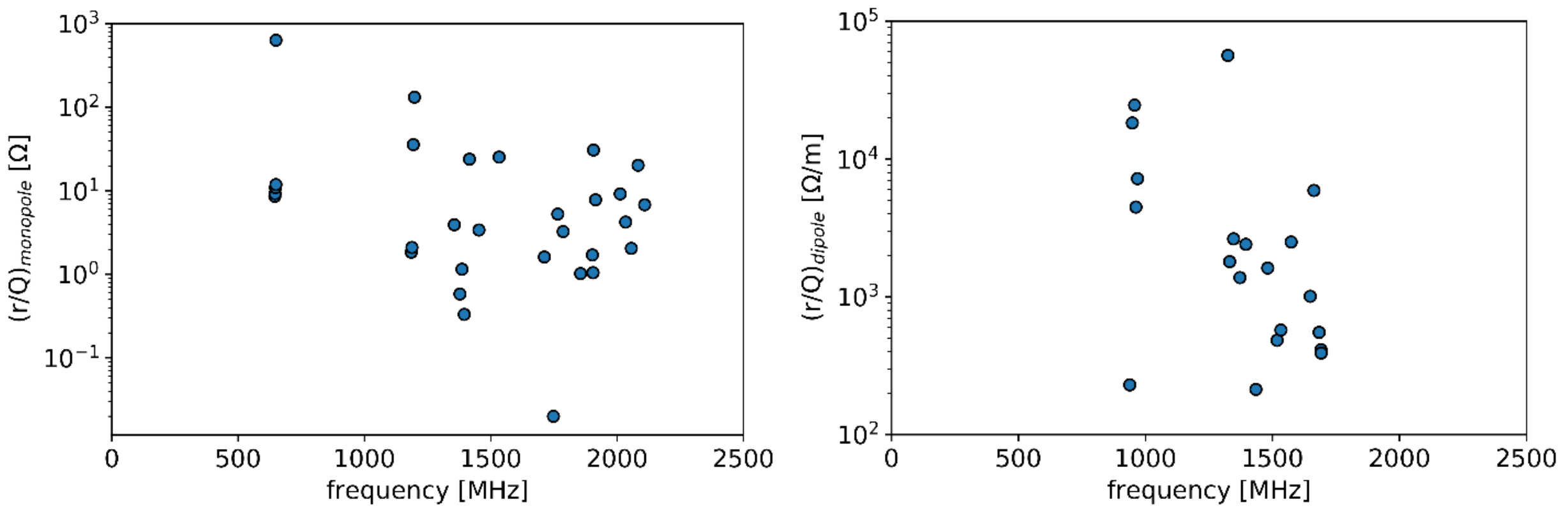

(a) (b)

*Figure 13: Impedances of (a) monopole HOMs and (b) dipole HOMs as a function of mode frequency.*

HOM Analysis Model

The present HOM model considers a continuous train of point-like bunches passing through the 5-cell cavity and accelerated by the cavity voltage. Bunch frequency is $f_b = 650$ MHz and each bunch has charge of $q_b = 154$ pC. Energy gain of bunches at the exit of the cavity is 10 MeV. Each passing bunch induces decelerating voltage in monopole mode $k$, given by:

$$U_k^{\parallel} = -\frac{1}{2}\left({R}/{Q}\right)_k \omega_k q_b \tag{2.1}$$

where $\omega_k = 2\pi f_k$ and $\left({R}/{Q}\right)_k$ are circular frequency and impedance of mode $k$. According to Wilson theorem of beam loading, the bunch "sees" half of its induced voltage. As the bunch passes through the cavity, the induced voltage evolves with time according to:

$$U_k^{\parallel}(t) = U_k^{\parallel}\exp\left[\left(j - \frac{1}{2Q_k}\right)\omega_k t\right] \tag{2.2}$$

where $j = \sqrt{-1}$. The total longitudinal voltage seen by bunch $N$ from the beginning of the bunch train is then given by:

$$V_N^{HOM,\parallel} = Re\sum_{k=1}^{P} U_k^{\parallel}\left\{\frac{1}{2} + \sum_{n=1}^{N-1}\exp\left[\left(j - \frac{1}{2Q_k}\right)\frac{\omega_k n}{f_b}\right]\right\} \tag{2.3}$$

where $P$ is total number of monopole modes, excluding the accelerating mode. If bunches have transverse displacement $x_b$ from the cavity axis, they also excite dipole modes. The induced transverse "kick" voltage is given by:

$$U_k^{\perp} = \frac{1}{2} j c q_b \left({R}/{Q}\right)_k x_b \tag{2.4}$$

where $c$ is speed of light. Total transverse kick voltage seen by bunch $N$ from $Q$ dipole modes is calculated as the following:

$$V_N^{HOM,\perp} = Re\sum_{k=1}^{Q}\sum_{n=1}^{N-1} U_k^{\perp}\exp\left[\left(j - \frac{1}{2Q_k}\right)\frac{\omega_k n}{f_b}\right] \tag{2.5}$$

Finally, the bunch transverse deflection angle as the ratio of kick voltage to total longitudinal momentum is given by:

$$X'_N = \frac{V_N^{HOM,\perp}}{(pc)_{\parallel}} \tag{2.6}$$

Figure 14 shows longitudinal voltage excited in monopole modes. Black markers show bunch-by-bunch voltage, green markers show cumulative mean voltage value, red markers show cumulative R.M.S. voltage value. The main contribution to longitudinal voltage is due to mode #4, 649.4 MHz. Quasi-periodic oscillations at 0.6 MHz, the frequency difference between bunch and mode frequencies. Maximum longitudinal voltage during transition is less than 1000 V, which is very small compared to cavity voltage 10 MV.

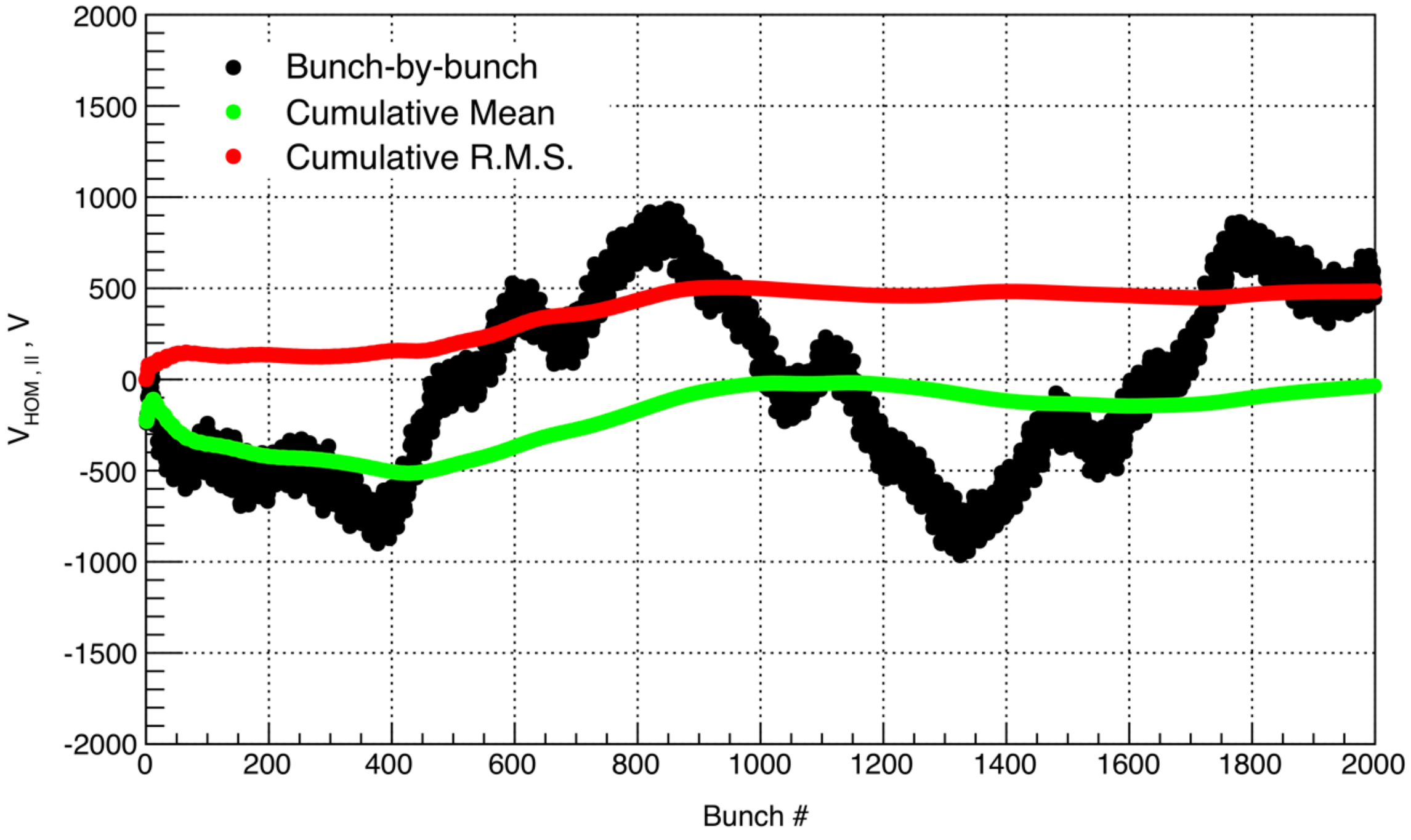

*Figure 14: Monopole HOM voltage.*

Results for dipole modes excited by bunches with 1 mm transverse displacement are shown in Figure 15. Quasi-periodic oscillation at 23.9 MHz (frequency difference between dipole mode #6 and 1st beam frequency harmonic). Maximum kick voltage from dipole modes during transition is -13 V. This is extremely small compared to longitudinal momentum 10 MeV. Corresponding transverse deviation angle is ~0.001 mrad.

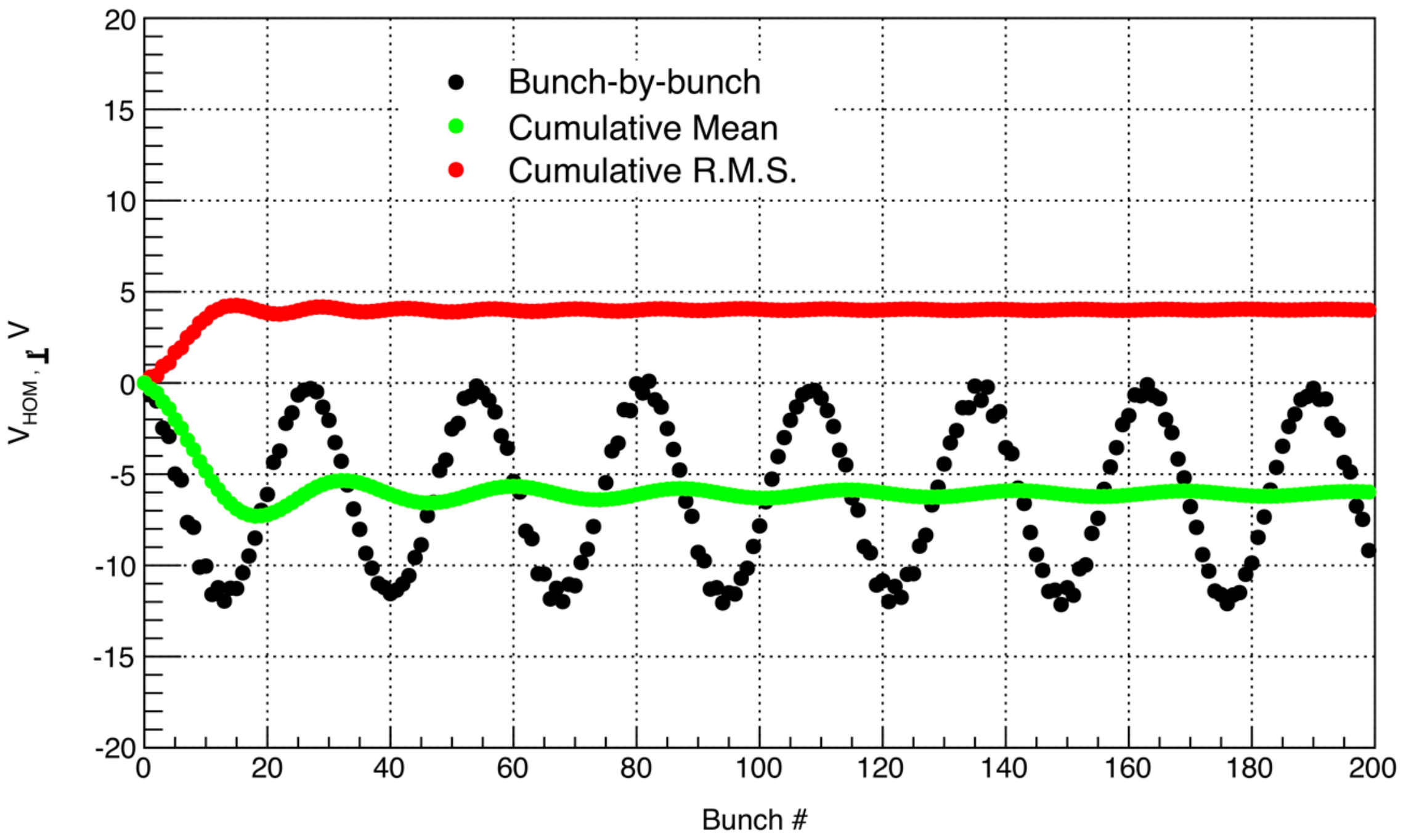

*Figure 15: Dipole HOM voltage.*

We note that both longitudinal and transverse voltage are proportional to the beam current. If we assume, that effects of HOMs should not exceed 0.1% on longitudinal and 0.1 mm on transverse beam dynamics, we still have at least a factor of 10 margin on beam current. Therefore, we conclude that HOMs are not expected to be an issue for the 5-cell cavity.

## d. Design of fundamental power coupler

RF power is fed to the 5-cell SRF cavity using two couplers, each sustaining 500 kW cw power with ~10% reflection. The coupler design is presented in this section, focusing on its RF performance, structural strength, and cryogenic loading to 4 K. The present design is motivated by the 650 MHz couplers developed and tested for the Proton Improvement Plan II (PIP-II) accelerator [17], which uses a ceramic window to separate the air side of the coupler, which receives power from an RF source, from the vacuum side, which delivers power to the cavity.

### *i. Ceramic window*

The ceramic window configuration as depicted in Figure 16 has 100 mm outer diameter, 25.4 mm inner diameter, and 7 mm thickness. The disc is made of alumina with loss tangent of $10^{-4}$ or less and is brazed with the outer and inner copper bushings. The outer copper bushing is brazed with a stainless-steel ring that connects to the outer conductor of the coupler.

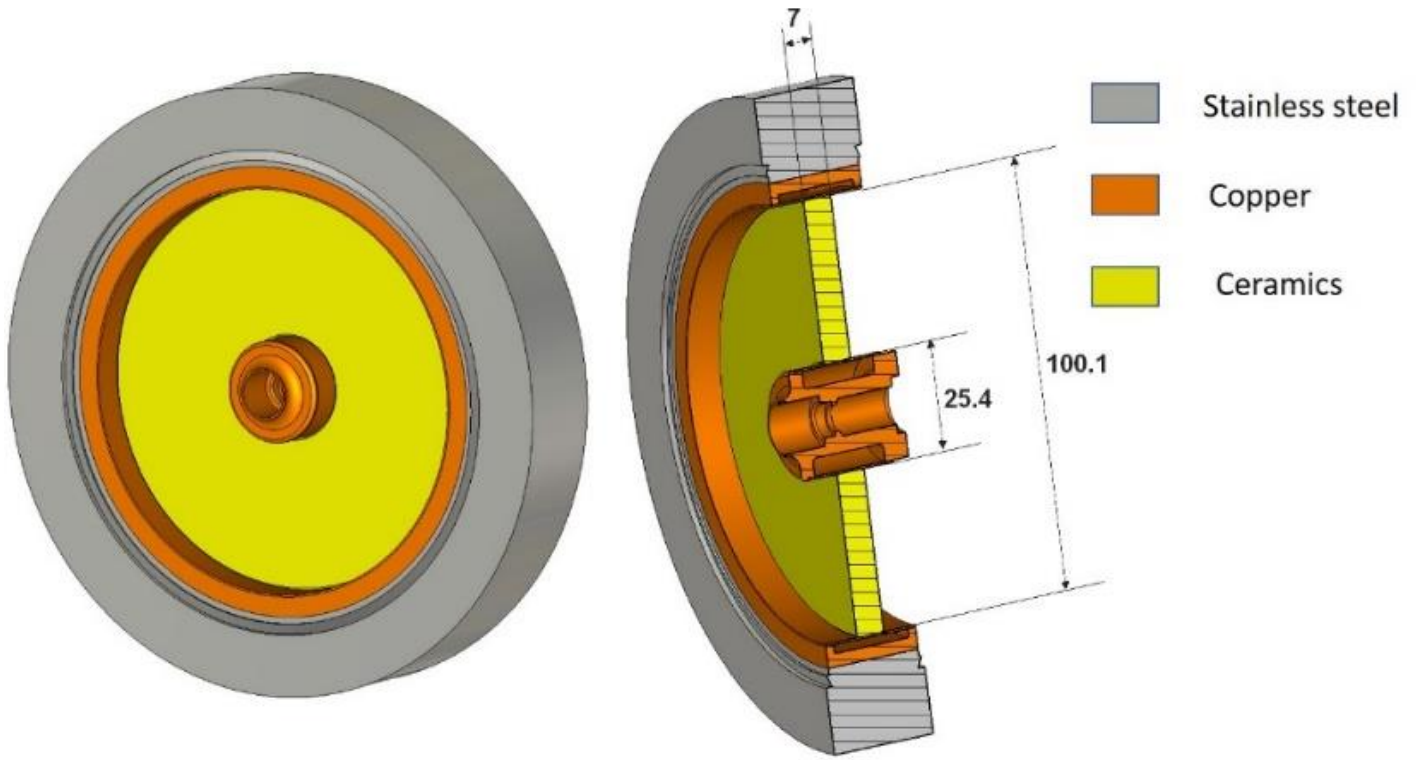

Figure 16: Configuration and dimensions of the coupler ceramic window.

*ii. Mechanical design of the power coupler*

The principal components of the coupler as assembled are depicted in Figure 17. To reduce static and dynamic cryogenic loading to 4 K, the coupler uses a copper electromagnetic shield (EMS) heat sunk only at ~50 K and no physical contact with the cavity. The EMS screens the stainless-steel outer conductor, aluminum gasket and stainless flange from electromagnetic field, thereby reducing ohmic losses in outer conductor, gasket, and flange. All RF losses are mostly concentrated in EMS and are intercepted by the 50K thermal intercept. The EMS also includes an iris which reduces thermal radiation from room temperature ceramic and protects the ceramic window from charged particles that can come from the cavity. The inner conductor of the coupler is a hollow copper channel cooled with forced flow of water. The calculated passband of the coupler is shown in Figure 18 with S31 being radiation through Kapton capacitor.

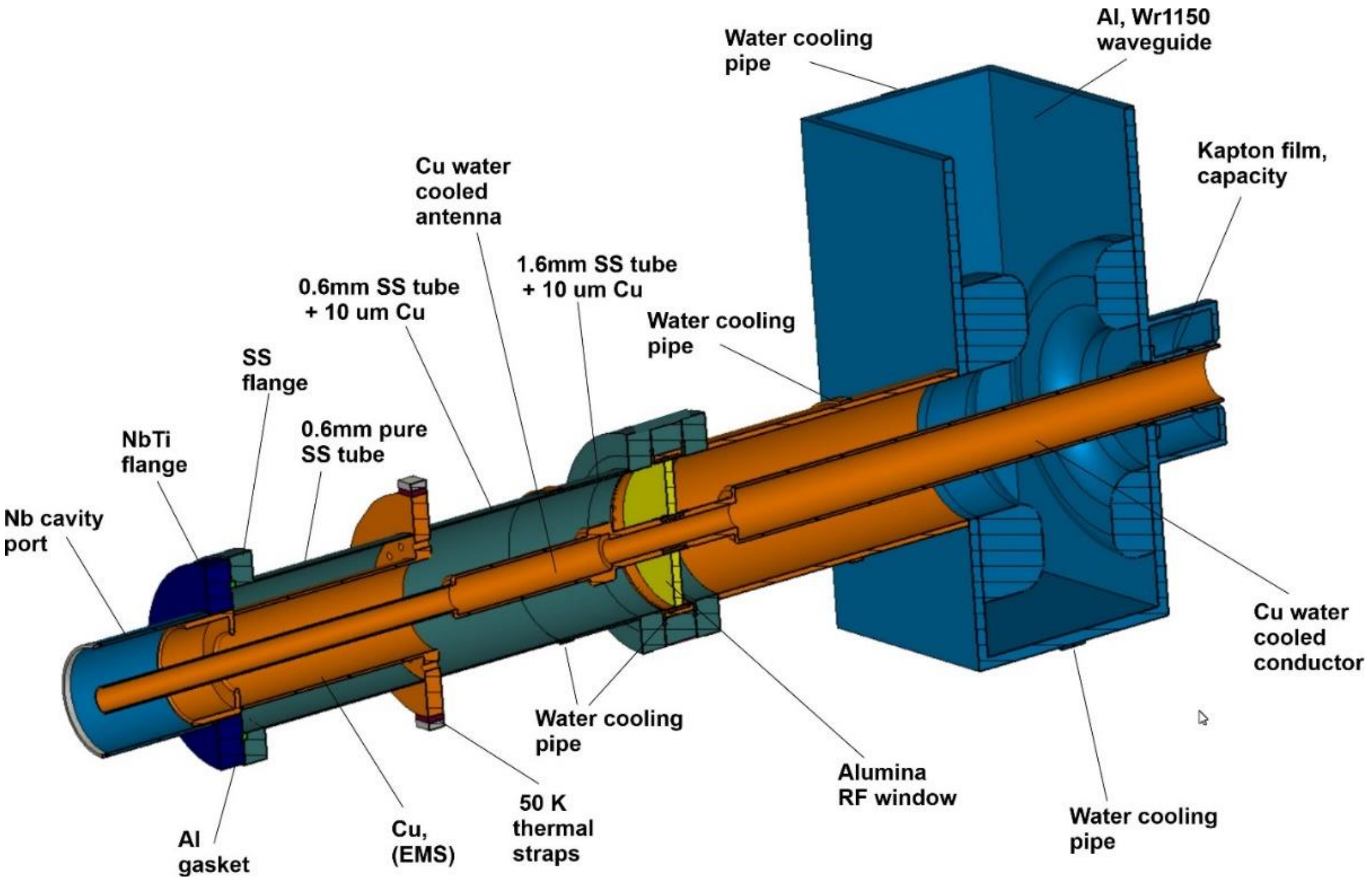

Figure 17: Cut view of configuration of 650 MHz, 500 kW, CW coupler.

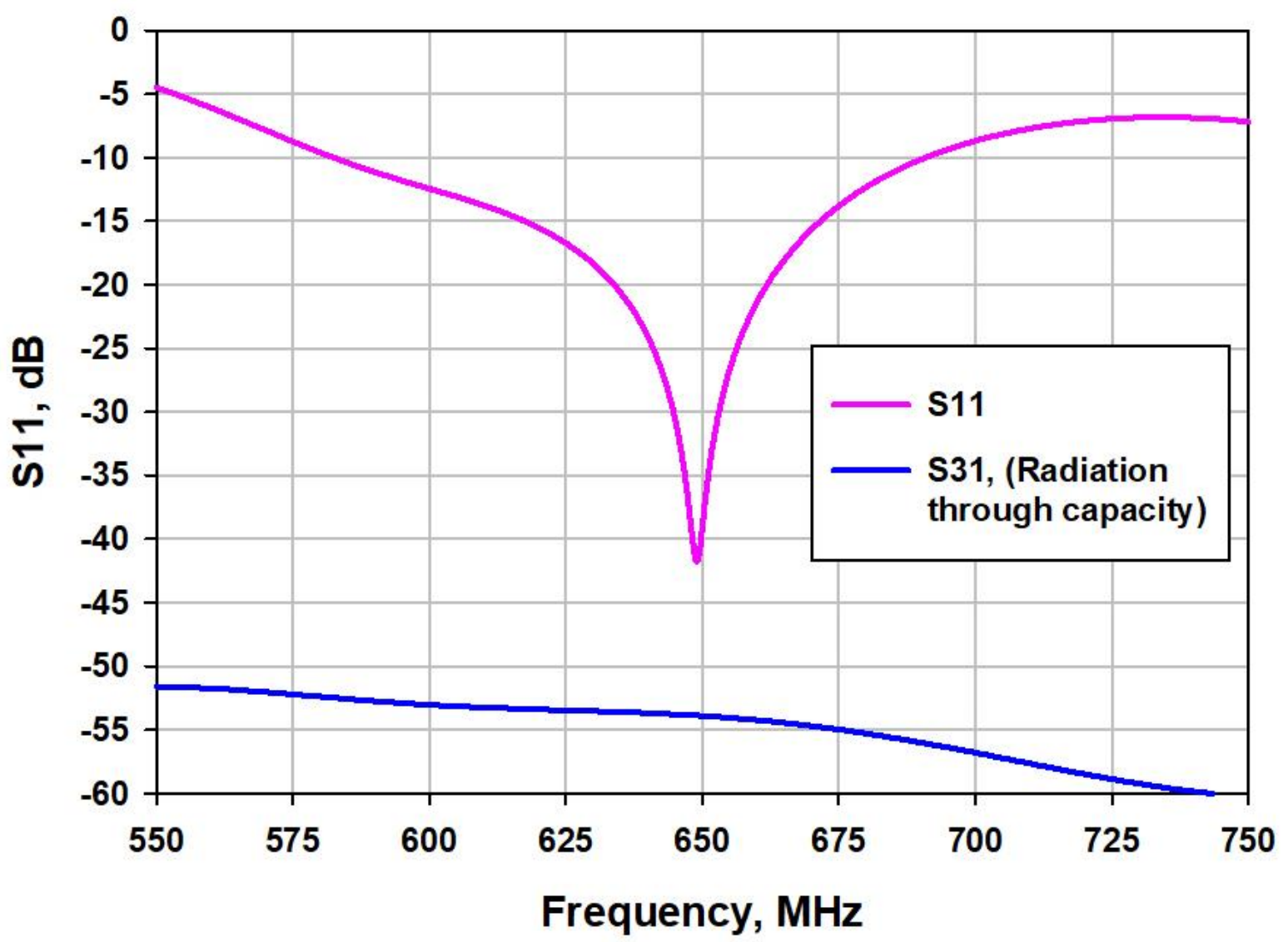

*Figure 18: Passband of the fundamental power coupler.*

*iii. Thermal analysis of the power coupler*

Thermal analysis is conducted to estimate cryo-loading (ohmic losses) in the coupler under steady operation with 500 kW forward propagation. The analysis uses the configuration and materials shown in Figure 19. The outer conductor is made of stainless steel with 0.6 mm wall thickness. The air side of the outer conductor (to the right of the 50 K intercept) is coated with 10 µm copper while the vacuum side (to the left of the 50 K intercept) is pure stainless steel, shielded by the EMS. A part of the EMS penetrates the cavity port made of superconducting niobium at ~4 K. There is a 0.8 mm gap between EMS and port wall to prevent thermal contact. All the EMS ohmic losses are intercepted at 50 K. The antenna, inner conductor, ceramic disc, and the air side of the outer conductor are cooled with forced flow of water at ~300 K. The outer copper sleeve of the ceramic window is cooled by water as well. The metal electrical conductivities and dielectric loss tangent of non-metals used in the thermal analysis are stated in Figure 19.

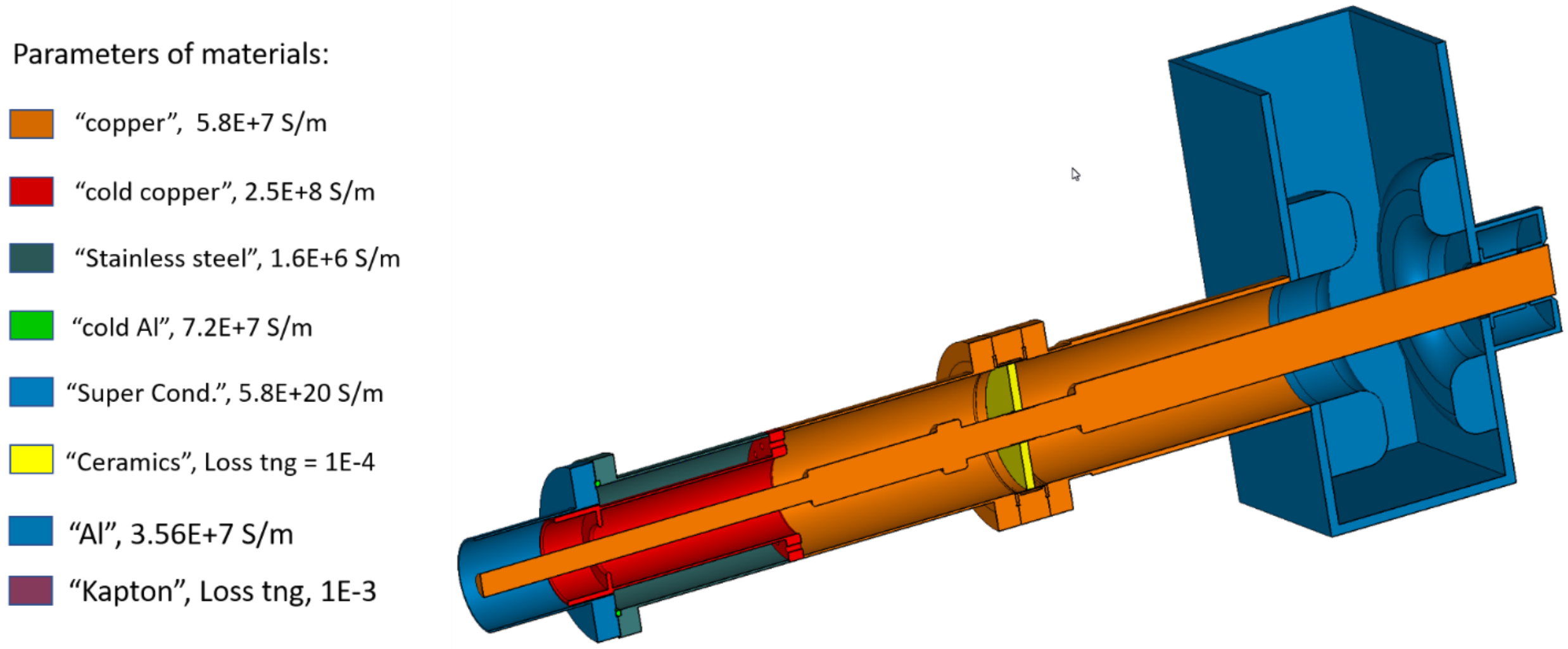

*Figure 19: Power coupler configuration for thermal analysis.*

Figure 20 shows the coupler temperature map at steady 500 kW of forward propagation with 10% reflection and the component-wise losses are listed in Table 3. With water cooling, the antenna tip is expected to operate at ~312 K. The conduction heat leak to the cavity port at 4 K is determined to be 0.6 W. In addition to this conductive heat leak, the cavity port will also experience incident radiation heat transfer from the warm EMS and the antenna tip. A finite element calculation of radiation heat transfer to the cavity coupler port using the temperature distribution of Figure 20 showed this additional load to be ~0.55 W. Therefore, the total 4 K heat load to the cavity port is expected to be ~1.2 W at full coupler forward power.

*Table 3: Losses in the coupler components for 500 kW forward propagation and 10% reflection.*

| **Component** | **Loss, W** | **Losses extracted by** |
|---|---|---|
| Outer conductor and flange to cavity port | 0.6 | Cavity port, ~4 K |
| Al gasket at the cavity port | $1.5 \times 10^{-3}$ | |
| EMS | 31.5 | Thermal intercept, ~50 K |
| Outer conductor (vacuum side), upstream of 50 K intercept | 29.5 | |
| Antenna | 570 | Cooling water, ~300 K |
| Ceramic disc | 43 | |
| Kapton | 5.5 | |
| Outer conductor (air side) | 44 | |
| Al waveguide | 255 | |

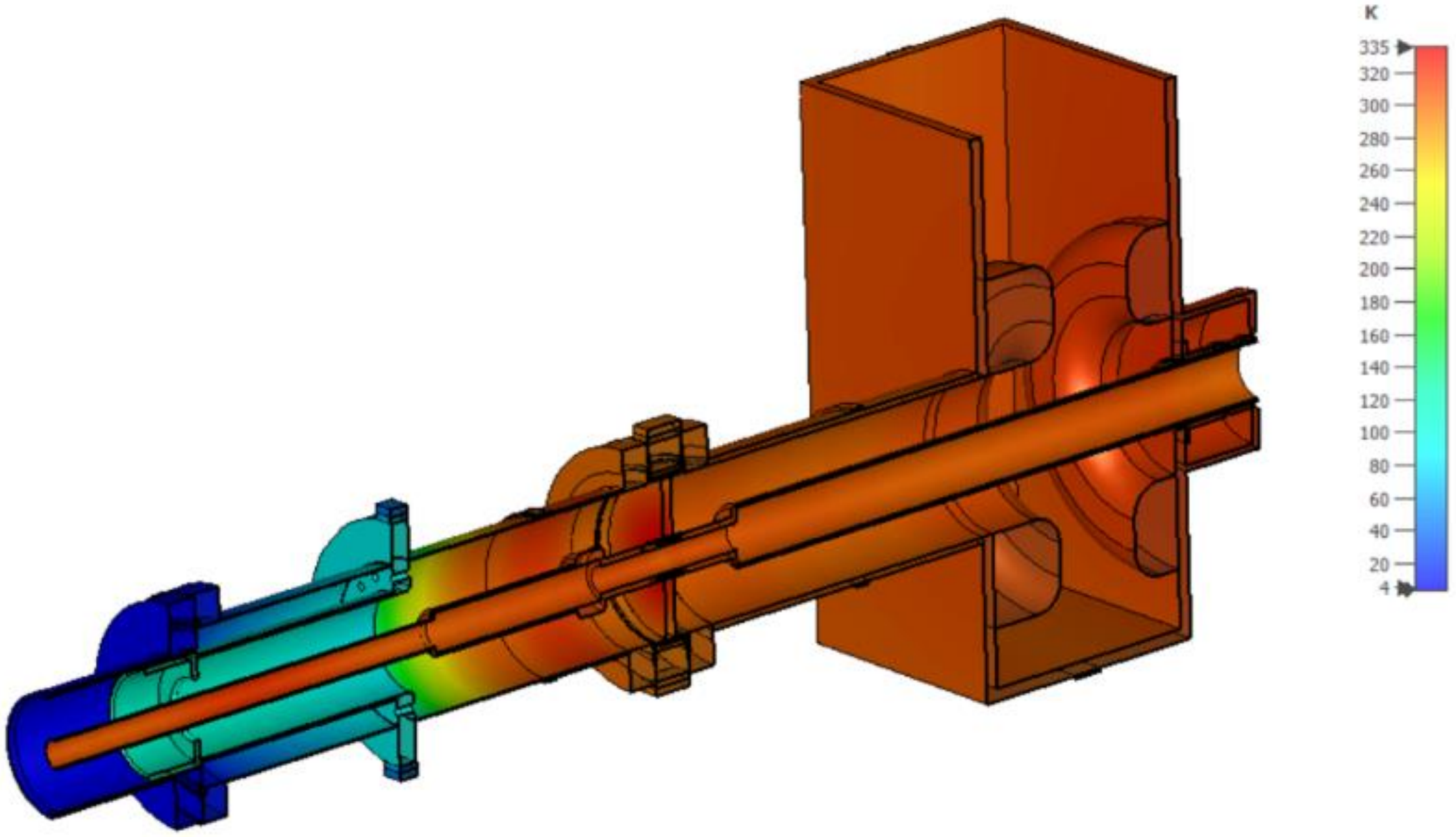

*Figure 20: Temperature distribution in the power coupler for 500 kW forward power with 10% reflection.*

*iv. Multipacting analysis of the power coupler*

The vacuum side of the power coupler is analyzed for multipacting issues, which are then mitigated by applying a DC bias. The vacuum side is divided for 4 sections and each section is simulated for multipacting with and without DC high voltage bias. The simulations are done for 866 kW, pure TW RF power, which is "field" equivalent of 500KW, 10% reflection. The four sections and graphs of rise of particles number *vs*. time are presented at Figure 21. Although multipacting is seen to exist in all the sections, it can be suppressed by applying a 5-6 kV DC bias.

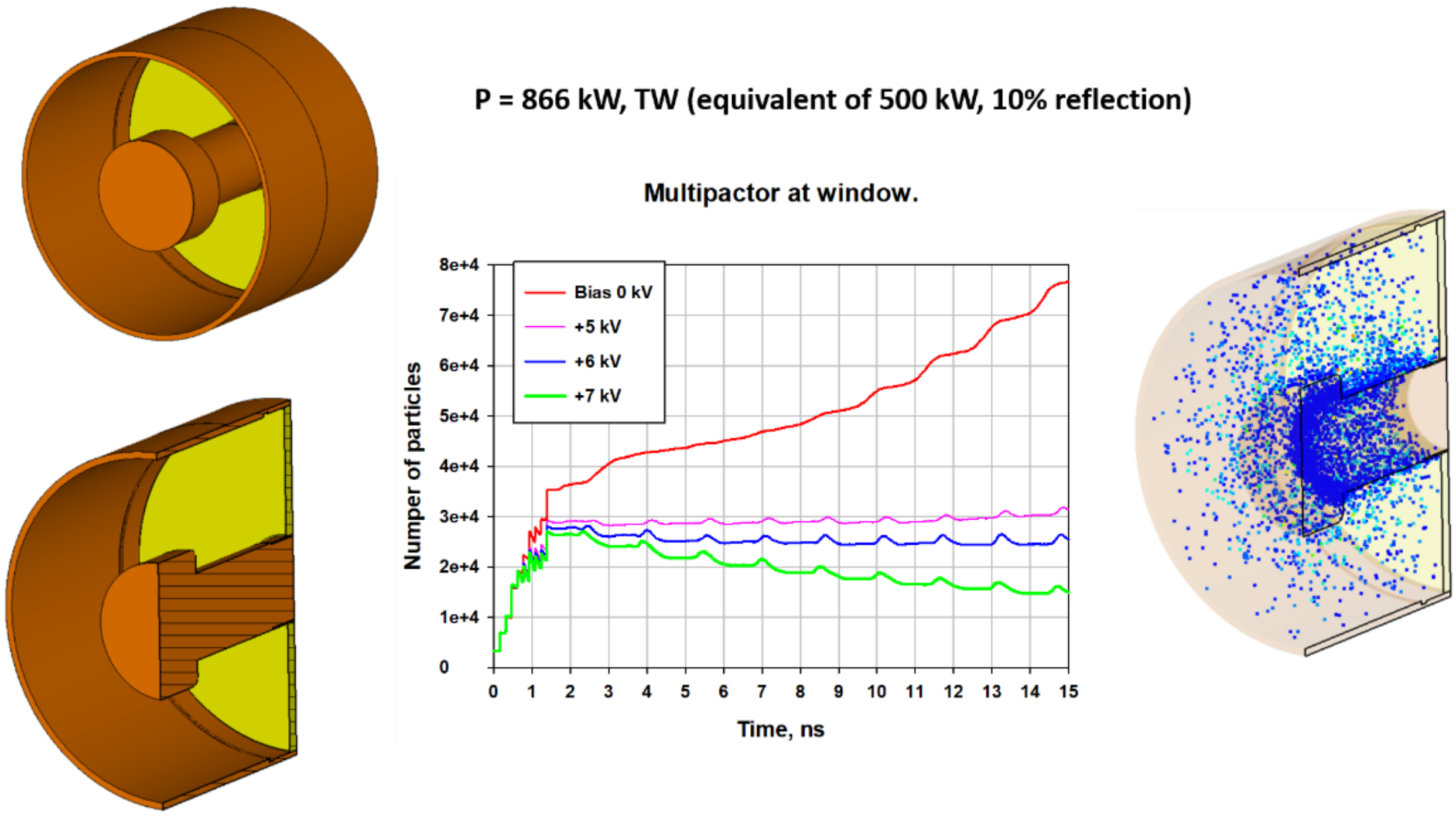

(a) Multipacting near the ceramic window

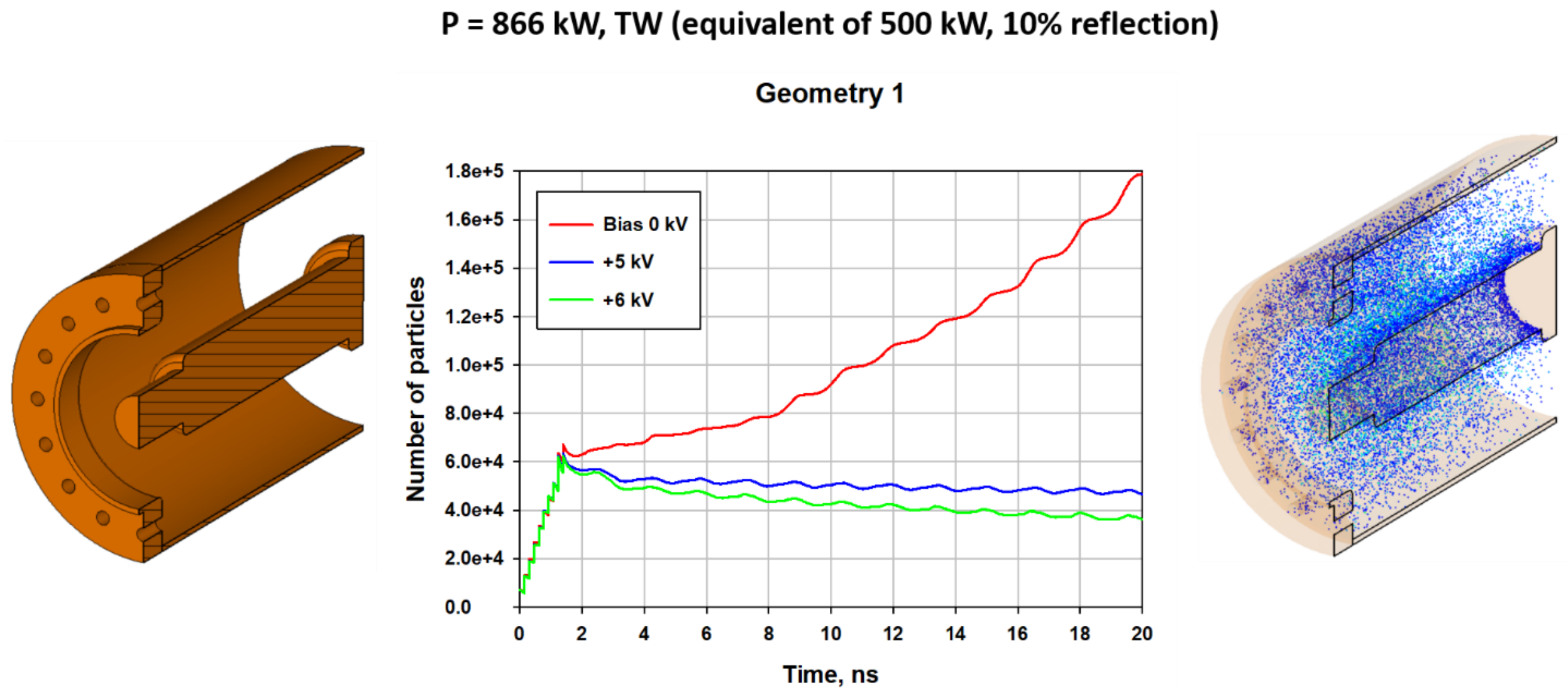

(b) Multipacting in the region near 50 K thermal intercept

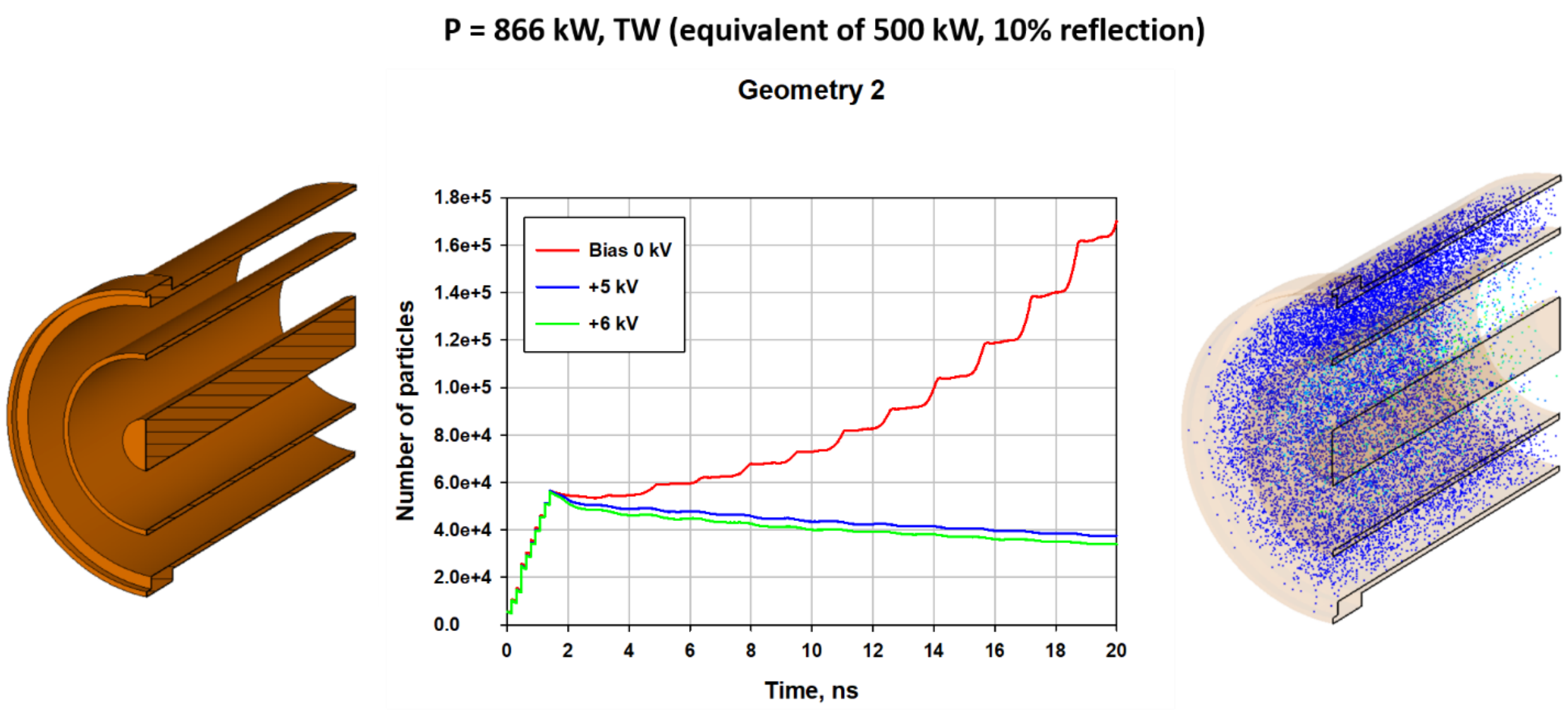

(b) Multipacting in the region of the EMS

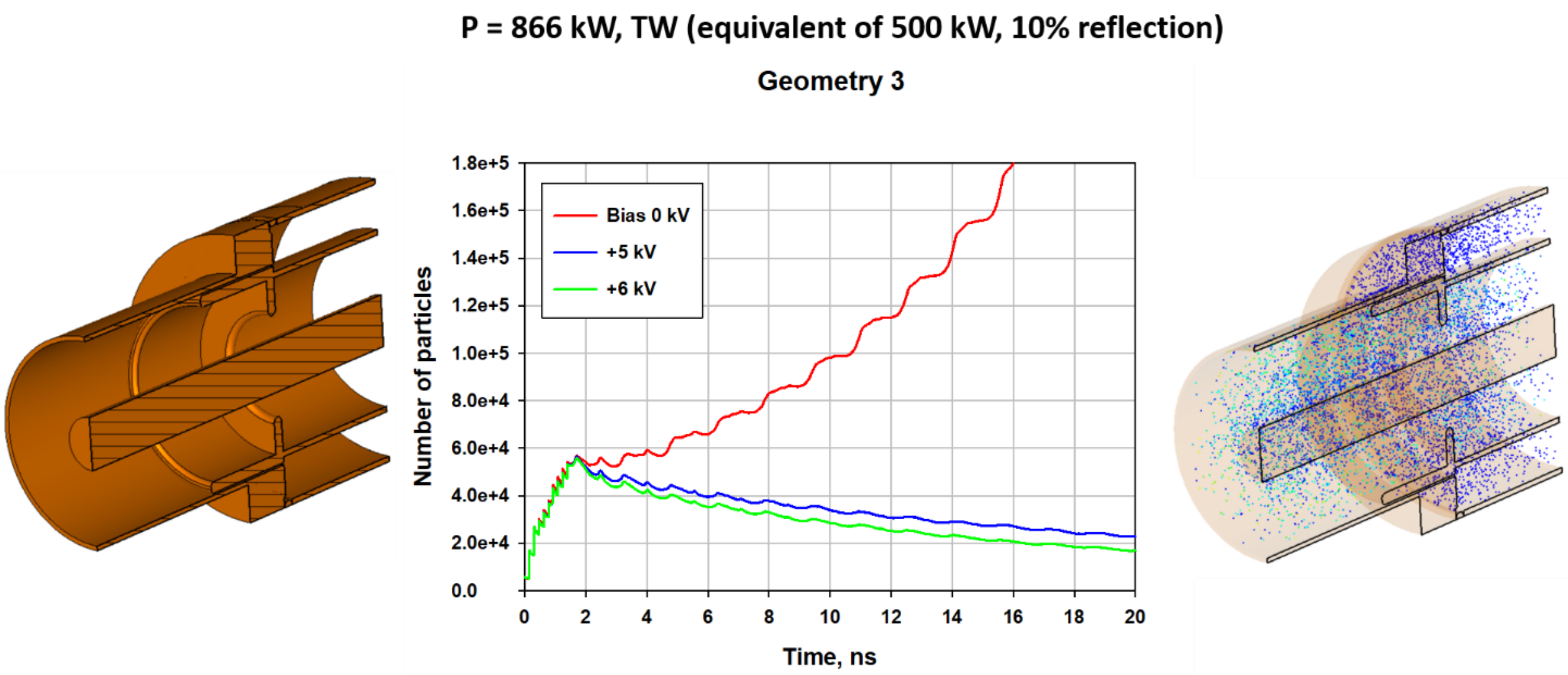

(d) Multipacting near the cavity port

*Figure 21: Simulated multipacting in the power coupler and suppression provide by a 5-6 kV DC bias.*

### *v. Summary of coupler design*

The main design parameters of the fundamental power coupler are summarized in the Table 4. With the use of the EMS, we find that the 4 K loading can be restricted to ~1.5 W at 500 kW of forward RF power. With two such power couplers, the total 4 K heat load is then expected to be ~3 W.

*Table 4: Summary of the power coupler design parameters.*

| Parameter | Value |
|---|---|
| Frequency | 650 MHz |
| Operating power | 500 kW, 10% ref. |
| Cryogenic loading to 4K | ~1.5 W |
| Cryogenic loading to 50K | ~62 W |
| Losses in ceramic disc | ~32 W |
| Maximum temperature at the ceramic window | ~335 K |
| Maximum temperature at the antenna tip | ~312 K |
| Bias voltage for supressing multipacting | 5-6 kV DC |

## e. Accelerator cryomodule design

A cross-section along the beamline of the cryomodule assembly is shown in Figure 22. The cryomodule includes a vacuum vessel, a 650MHz $Nb_3Sn$ cavity, eight two-stage cryocoolers (four on each side of the beamline as seen in Figure 22), and single-layer thermal and magnetic shields. The cavity is conduction-cooled to the cryocoolers as will be described later in this section. The thermal shield insulates the SRF cavity from ambient thermal radiations and intercepts the heat transmitted through the RF couplers and the beamline ports. The thermal shield is connected to the 1st cooling stages of the cryocoolers using a set of thermal straps [18] visible in Figure 22. Two types of cryocoolers are selected: six Cryomech PT420 offering a cooling capacity of 2.0 W at 4.2 K and two additional Cryomech PT425 with a higher cooling capacity of 2.5 W at 4.2 K located above the RF couplers. Strong magnetic fields can impair the intrinsic quality factor of the cavity, thereby reducing the attainable accelerating gradient for a given cryocooling capacity. A magnetic shield is provided to limit the total magnetic field on the surface of the SRF cavity to <10 mG. The magnetic shield is operated at room temperature to avoid additional cryogenic loading of the cryocoolers. The cold mass and the magnetic shield are all enclosed in a 1.95 m-long vacuum vessel. The total mass of the fully-assembled cryomodule is estimated to be 1750 kg. Design and analysis of the various cryomodule components are presented in the following sections.

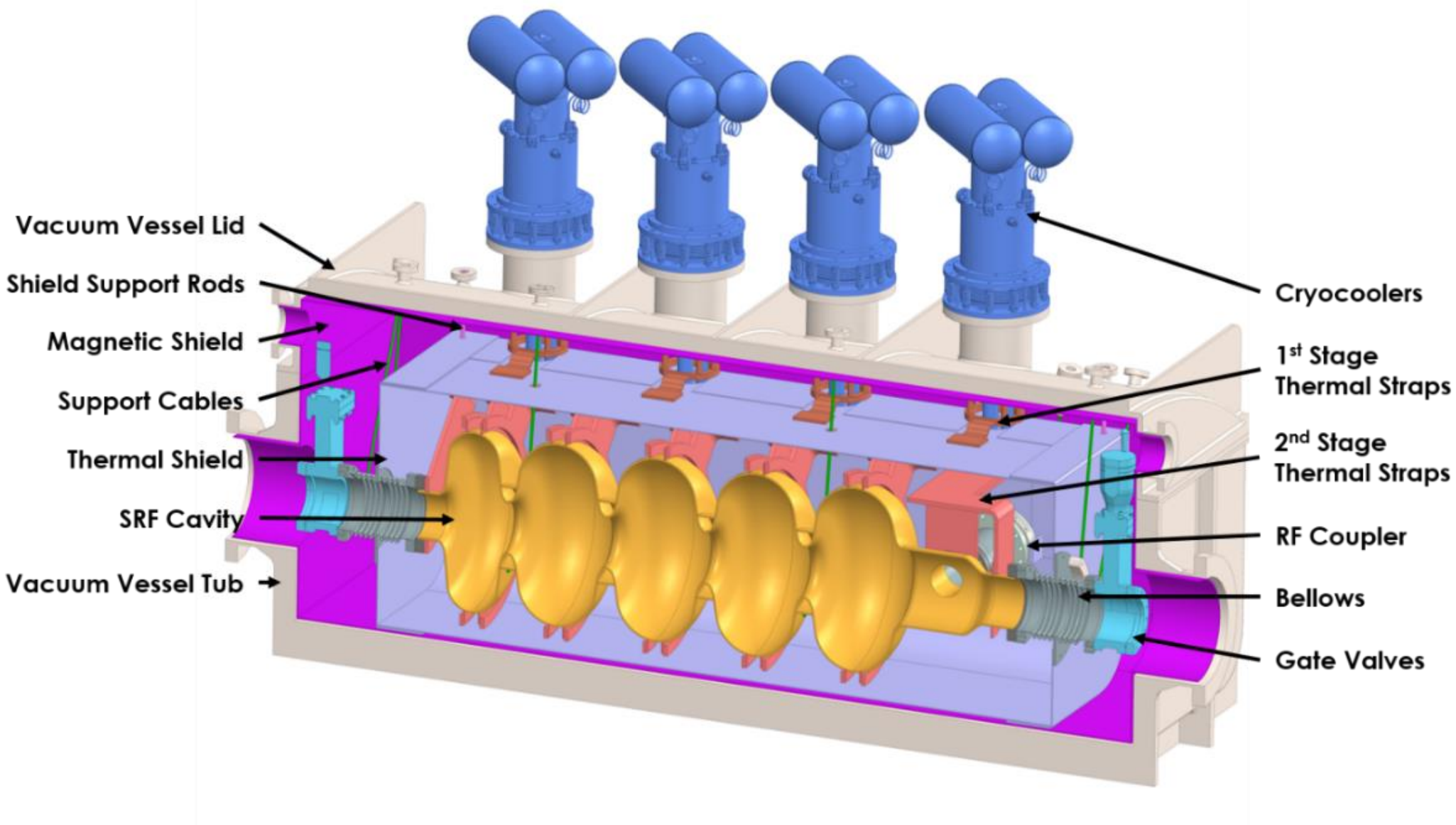

*Figure 22: Cross-section view of the SRF cryomodule assembly.*

### *i. SRF cavity cooling design and thermal analysis*

<u>Estimation of SRF cavity cryoloading</u>

Table 5 presents the calculated ~4 K heat load on the cavity and the details of the cryocoolers chosen to provide the required cooling load. The cavity is divided into two sections to simplify the heat load estimation – (1) the main body comprising the five elliptical cells and the inlet beam tube, and (2) the outlet side made of the two coupler ports and outlet beam pipe. The static heat leak contributions of thermal radiation from the thermal shield and *via* beam pipes, thermal conduction *via* cavity supports, beam pipes, and coupler ports are considered. The dynamic loading comprises beam loss, cavity RF heating, and coupler loading. Although not explicitly determined in this study we assume beam loss of 1 W, which is 1 ppm of the 1 MW average beam power. The cavity RF heating is initially estimated using 20 nΩ surface resistance for the $Nb_3Sn$ RF surface at 650 MHz and 10 MV voltage gain over the 5-cell cavity length (see Table 2 for the expression of dissipated RF power). The RF heat load will be revised in a subsequent section by accounting the temperature dependence of $Nb_3Sn$ surface resistance. While the total heat load from the couplers to the cavity is previously estimated to be ~3 W (~1.5 W per coupler, see Table 4), we take a conservative value of 3 W *per coupler* as seen in practice with a 500 kW coupler tested at Brookhaven National Laboratory [19].

In summary, the total cavity heat load is estimated to be 19.5 W. The cavity body and inlet side experiences 14 W of heat load, which can be extracted using six Cryomech PT420 cryocoolers operating at 4.45 K. The cavity outlet side has 6.5 W of cryoloading, which is manageable using two Cryomech PT425 cryocoolers operating at 4.6 K.

*Table 5: Calculated heat load on the 5-cell SRF cavity.*

| **Component** | **Heat load [W]** | | **Cryocooler selection** |
|---|---|---|---|
| Cavity body | RF dissipation @ 10 MeV and Rs = 20 nΩ | 12.5 | |
| | Radiation from thermal shield | 0.05 | |
| | Beam loss (assumed $10^{-6}$ parts of 1 MW average beam power) | 1 | |
| | Conduction through supports | 0.1 | |
| Cavity input side | Conduction through inlet beampipe | 0.05 | |
| | Radiation through inlet port [20] | 0.24 | |
| **Cavity body + inlet side** | | **14** | **6 x Cryomech PT420, cooling capacity ~14.3 W @ 4.45 K** |
| Cavity output side | Coupler [19] | 6 | |
| | Conduction through outlet beampipe | 0.05 | |
| | Radiation through outlet beampipe [20] | 0.24 | |
| **Cavity outlet side** | | **6.3** | **2 x Cryomech PT425 cryocoolers, cooling capacity ~6.5 W @ 4.6 K** |
| **Total on cavity** | | **19.5** | |

SRF cavity cooling design and analysis

The 5-cell cavity shown in Figure 23(a) is made of a 4 mm thick niobium shell (SRF grade, RRR>300) around the profile given in Figure 7. The cavity inner surface is coated with a ~2 µm thick layer of $Nb_3Sn$, which enables low dissipation operation at ~4 K temperature. Each cell has two 4 mm thick conduction cooling rings made of SRF grade niobium that are e-beam welded at about 12.5 mm on either side of the cell's equator. The two coupler pipes as well as the inlet and outlet beampipe carry thermal intercepts made of niobium. All the port flanges are made of niobium-titanium alloy and are e-beam welded to the beam and coupler ports.

High purity (5N or >99.999% pure) aluminum thermal links are used to conductively connect the 5-cell niobium cavity with the cryocoolers. Two separate thermal links are used – one for cooling the inlet beam pipe and the five cells, and the second for cooling the outlet beam pipe and coupler pipes. The link components are cut out of 4 mm or 6.35 mm thick sheets of commercial 5N aluminum and then bent into final shapes. The components are then connected to each other and to the cavity cell cooling rings, resulting in the configuration depicted in Figure 23(b). All the connections within the link as well as to the cavity are made using off-the-shelf nuts, bolts, and disc springs that enable easy disconnection if required. Although not shown in Figure 23 to avoid clutter, the niobium rings on the cavity cells carry several bolt holes to connect with the aluminum thermal link components.

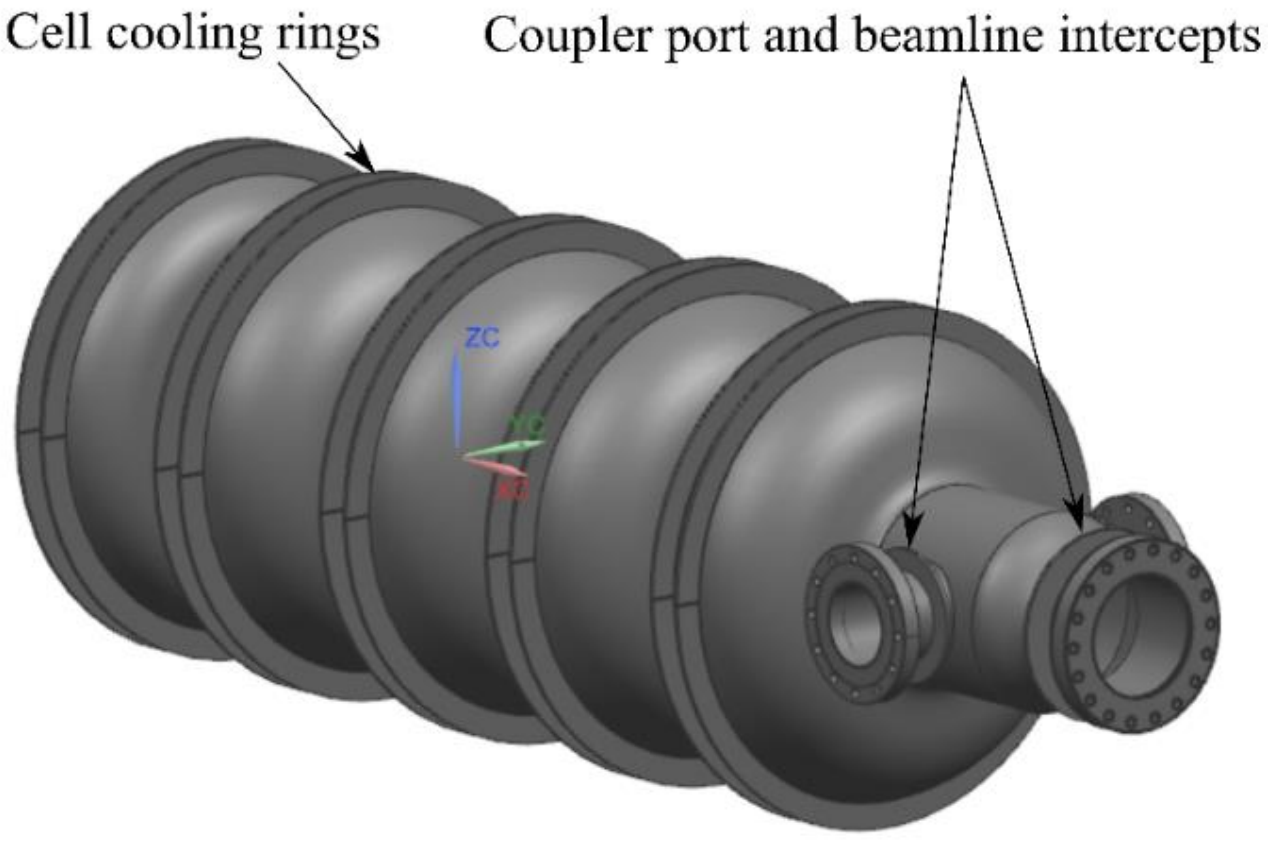

(a)

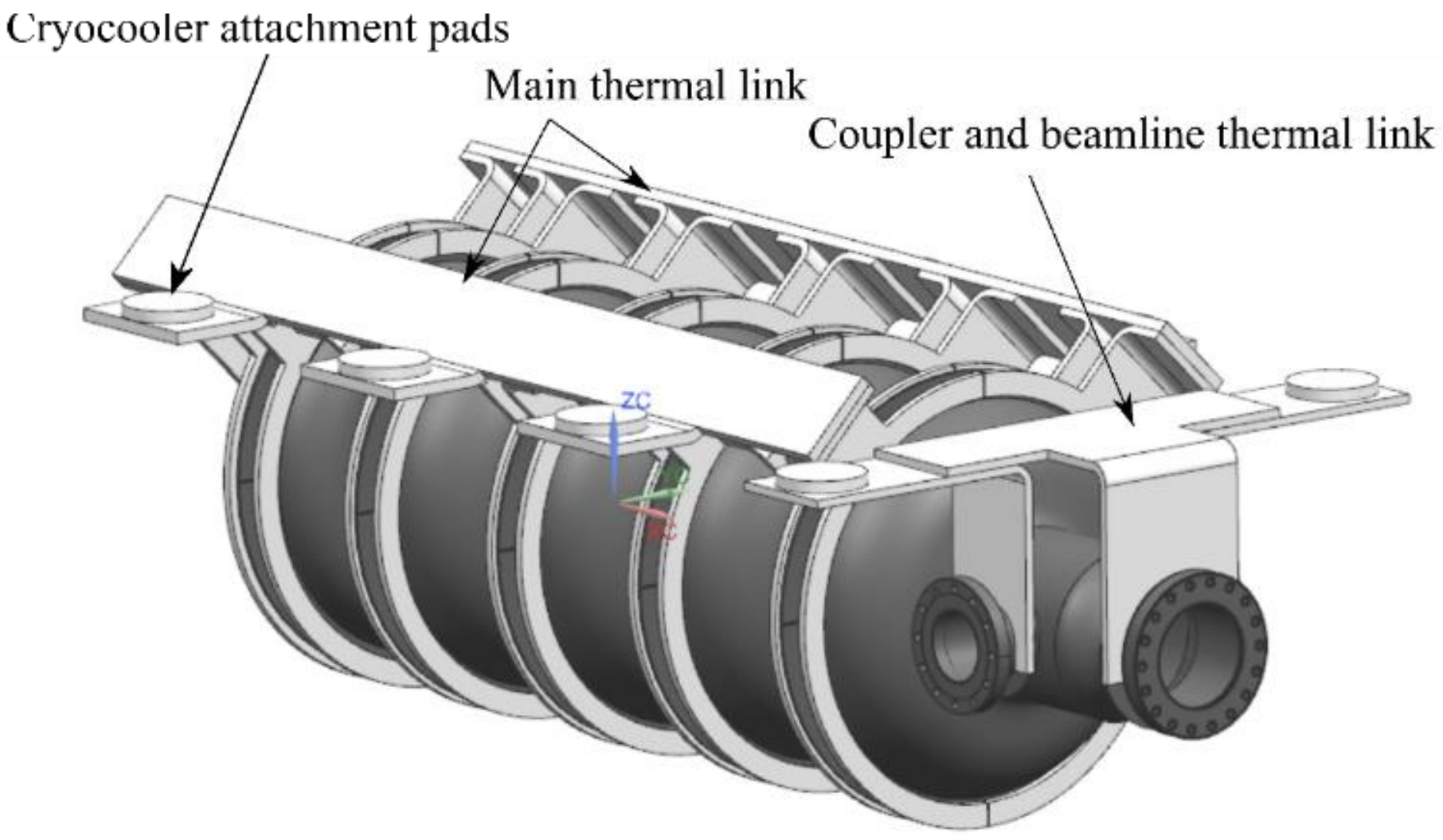

(b)

*Figure 23: (a) Rendering of the 5-cell cavity with e-beam welded cooling rings and (b) thermal links attached to the 5-cell cavity by bolting.*

The effectiveness of the aluminum thermal links is evaluated by systematic finite element simulations. The goals here are (1) to obtain reasonably small temperature drop between the cavity RF surface and the cryocoolers and (2) to obtain reasonably uniform surface temperature of the cavity. The simulations use the following two heat transfer boundary conditions: (1) all heat flows at appropriate locations on the cavity as listed in Table 5 and (2) temperature dependent cooling capacity of the cryocoolers (measured in-house), imposed on the cryocooler attachment pads. The simulations use temperature dependent $Nb_3Sn$ surface resistance (taken as the sum of 20 nΩ residual and BCS resistance calculated using SRIMP [21]), temperature dependent thermal conductivities of 5N aluminum [22] and niobium [23], and thermal contact resistance across the bolted connections [22,24]. Figure 24 shows the steady state temperature profile of the cavity – thermal link assembly at 10 MeV, 100 mA accelerator operation as well as a temperature line graph along the cavity arc length from the inlet to the outlet. The simulated temperature profile shows the maximum $T_{iris} – T_{equator}$ <0.25 K and $T_{cell} – T_{cryocooler}$ < 0.5 K.

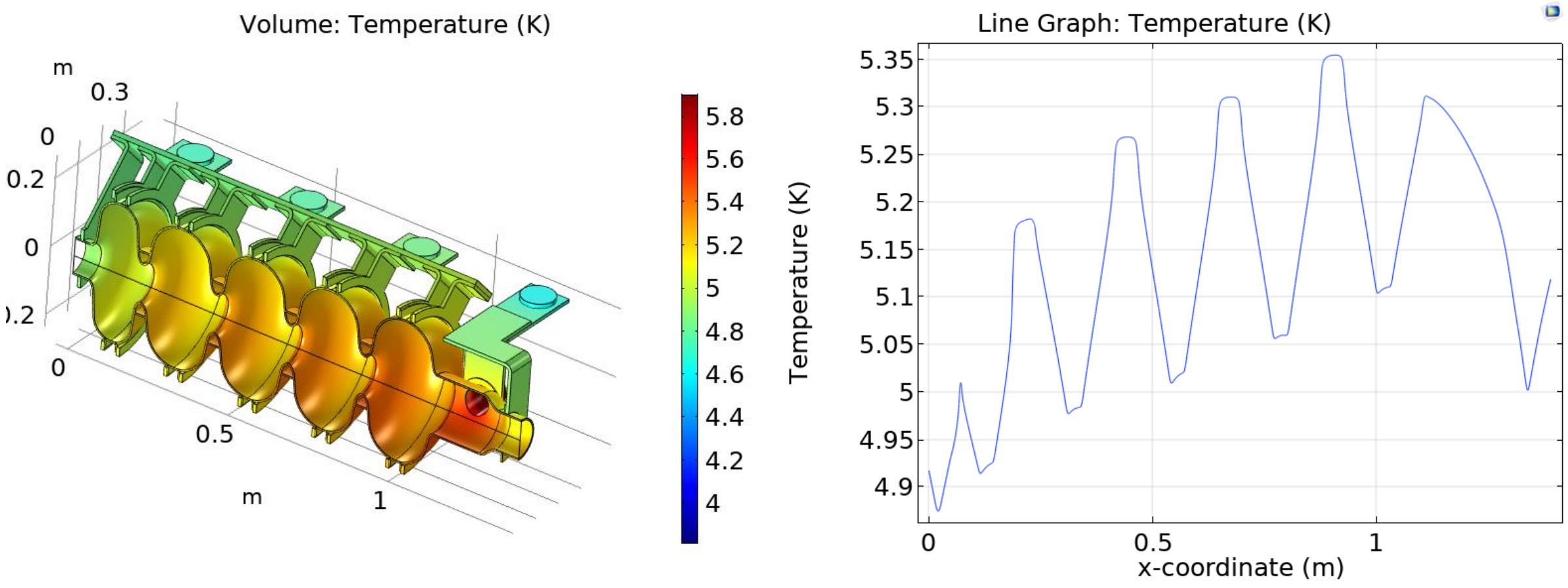

*Figure 24: Steady state cavity temperature profile for 10 MeV, 100 mA operation – surface temperature map (left) and line graph along the cavity wall profile (right).*

The conduction cooling technique presented above has been experimentally validated by the present authors as reported in their prior work [25,26,27]. Herein, a single-cell 650 MHz Nb3Sn cavity was conduction-cooled using a Cryomech PT420 cryocooler. The single-cell cavity produced 10 MV/m cw accelerating gradient over 0.23 m length. The same gradient on the present 5-cell cavity is equivalent to >10 MV voltage gain.

*ii. Design and analysis of other cryomodule components*

<u>Thermal shield</u>

The thermal shield performance is evaluated using a heat transfer analysis in COMSOL Multiphysics. The COMSOL thermal model is presented in Figure 25(a). The thermal shield, made of 2.5 mm thick aluminum panels, is connected to the first stages of the cryocoolers using copper-based thermal joints. The cryocooolers ensure the thermal shield temperature stays below 50 K, as demonstrated below. Openings in the shield panels allow for the cryocooler heads, the support straps, the power couplers, and the beamline to be connected to the SRF cavity. In order to reduce conductive heat transfer with the surroundings, the thermal shield is suspended by a set of titanium-64 rods hanging from the top plate of the vacuum vessel.

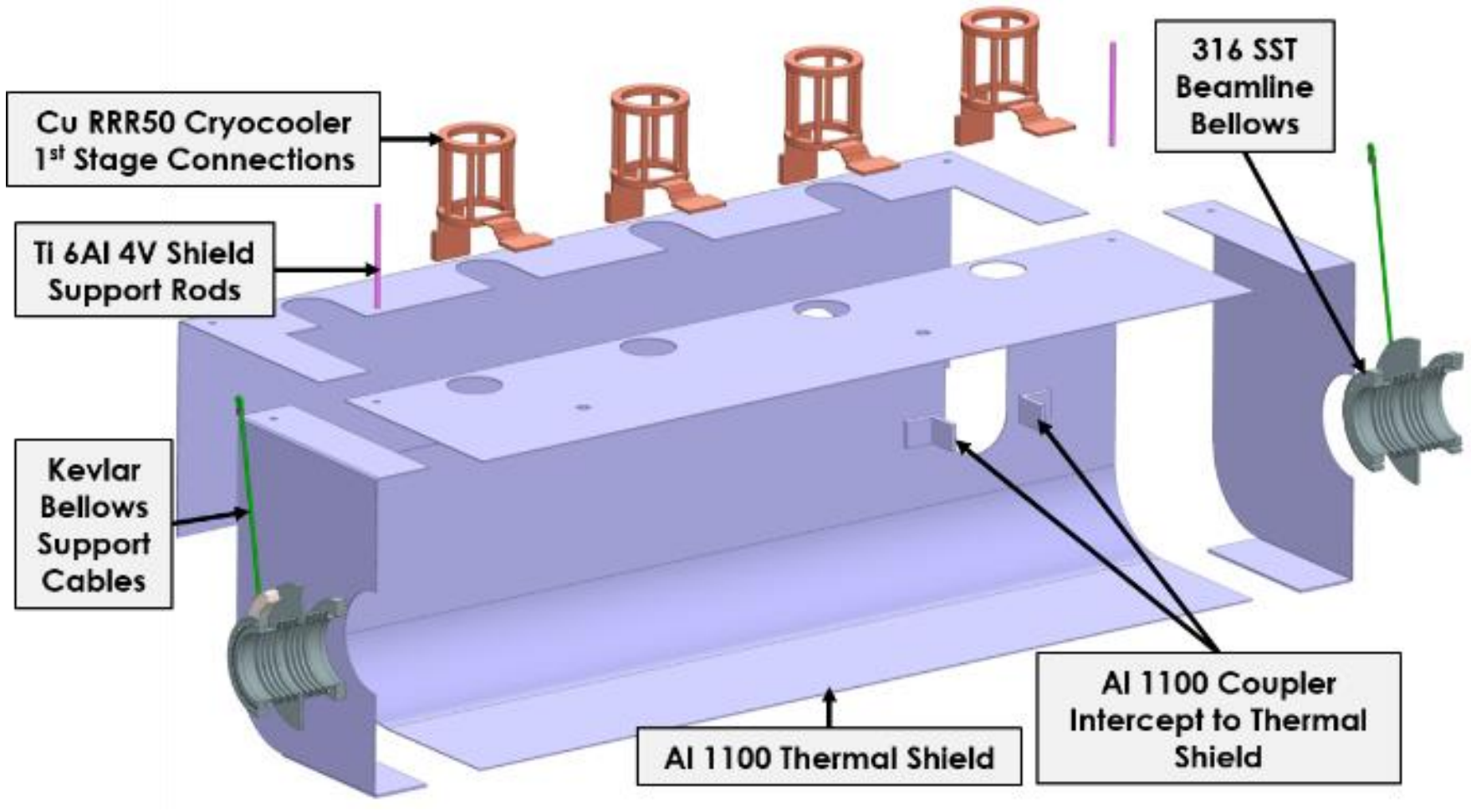

(a)

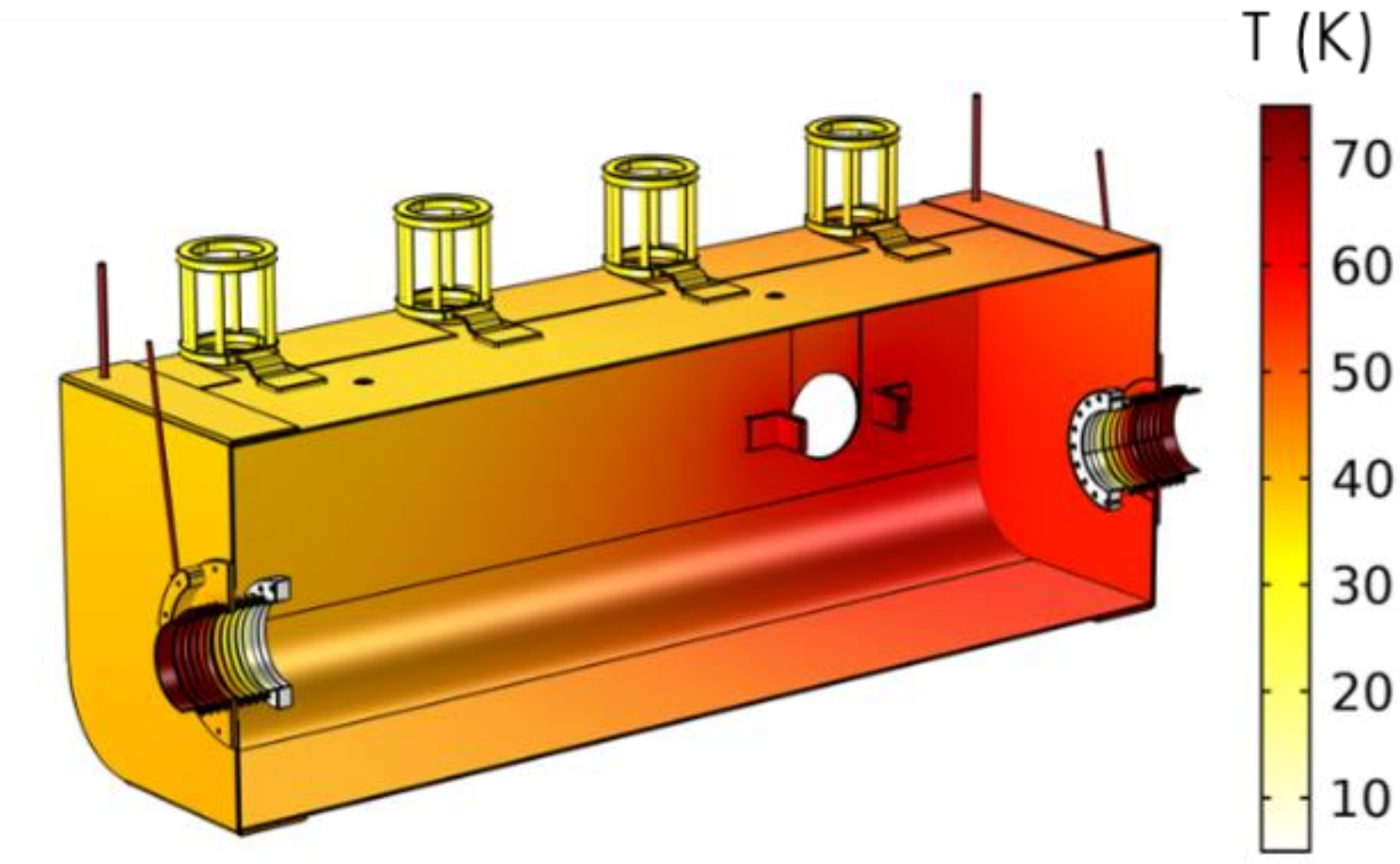

(b)

*Figure 25: (a) Thermal shield exploded view showing how the shield panels and components are assembled (b) Thermal shield temperature distribution during steady state operation of the accelerator.*

The material thermal properties are modeled using temperature-dependent thermal conductivity data [28]. Steady-state thermal boundary conditions are imposed as following:

- Each RF coupler introduces 60 W of heat flux into the shield, according to the results in Table 3.
- A thermal radiative heat flux of 1.5 W/m$^2$, estimated for an average thermal shield temperature of 50 K and an outer temperature of 300 K, is imposed on all outward facing surfaces. Thermal radiation from the thermal shield at 50 K to the cavity at ~5 K is estimated to be 50 mW. Accounting for these two radiation heat flows, the net radiative heat flux incident on the thermal shield is approximately 8 W.
- The cavity connections to the beamline bellows are assumed at a fixed temperature of 5 K.
- The ambient temperature ends of the beamline bellows, titanium rods, and kevlar support cables are set at 300 K.
- Heat flux through the cryocooler 1$^{st}$ stages is derived empirically as Q[W] = 117.9 – 3.93T[K] for PT420 cyrocoolers and Q[W] = 147.4 – 4.91T[K] for PT425 cyrocoolers [29].

In Figure 25(b), the thermal shield temperature varies from a minimum of 32 K at the interfaces with the cryocoolers, to 39 K and 51 K at the joints with the two beamline bellows, and to a maximum of 63 K at the connections with the RF couplers. The average temperature of the thermal shield is 43 K. The cryocoolers extract a total of 137.2 W from the thermal shield, which is within the operating range of the cryoccoler's 1$^{st}$ cooling stages. All other sources of heat, listed in Table 6, are comparatively minimal. The model predicts that only 0.41 W is transferred to the cavity through the beamline. Therefore, the thermal shield effectively isolates the cavity from the beamline at room temperature.

*Table 6: Thermal balance across the thermal shield assembly.*

| **Heat Input** | | **Heat Output**(*) | |
|---|---|---|---|
| *Conduction through RF coupler intercepts* | *120.0 W* | *Cryocoolers Pair#1 (Cryomech PT420)* | *23.1 W* |
| *Thermal radiations with surroundings* | *8.0 W* | *Cryocoolers Pair#2 (Cryomech PT420)* | *25.8 W* |
| *Conduction through threaded rod supports* | *2.25 W* | *Cryocoolers Pair#3 (Cryomech PT420)* | *35.5 W* |
| *Conduction through Kevlar support cables* | *0.25 W* | *Cryocoolers Pair#4 (Cryomech PT425)* | *52.8 W* |
| *Conduction through beamline bellows* | *7.2 W* | *SRF Cavity* | *0.5 W* |
| ***TOTAL HEAT INPUT*** | ***137.7 W*** | ***TOTAL HEAT OUTPUT*** | ***137.7 W*** |

(*) *Cryocoolers are labelled from the closest pair (Pair#1) to the beamline input to the closest pair (Pair#4) to the beamline output.*

Magnetic shield

The shielding material, MuMetal, is modeled with a constant permeability relative to vacuum of 30,000 (to be conservative, this value is chosen to be 2.5 x smaller than the vendor specified value of 75000).

Magnetostatic simulations performed on the configuration shown in Figure 26(a) using COMSOL Multiphysics demonstrate that a single-layer magnetic shield operating near ambient temperature between the vacuum vessel and thermal shield can adequately provide for the <10 mG target background field at the cavity surface. The simulations are performed in the local Earth's magnetic field at Fermilab (Kane County, Illinois), which has a magnitude of 534 mG with components of 193 mG, 12 mG, 498 mG in the North and West directions and vertically toward the Earth's center, respectively [30]. The magnetic shield is modeled inside a spherical background domain of 8 meters in diameter, large enough for the boundaries to not be disturbed by the presence of the magnetic shield at the center. Simulations were compared with the beamline oriented at different angles with respect to the North, namely 0 deg, 45 deg and 90 deg. As shown in Figure 26, for all three cases, the magnetic flux density along the surface of the SRF cavity is fairly uniform and does not exceed 10 mG. The peak value of the total flux at the cavity surface is 9.6 mG when the beamline is oriented at a 45 deg with respect to North and 9 mG for the orientations at 0 deg and 90 deg. Therefore, the operation of the accelerator is expected to not be affected by its orientation with respect to the North.

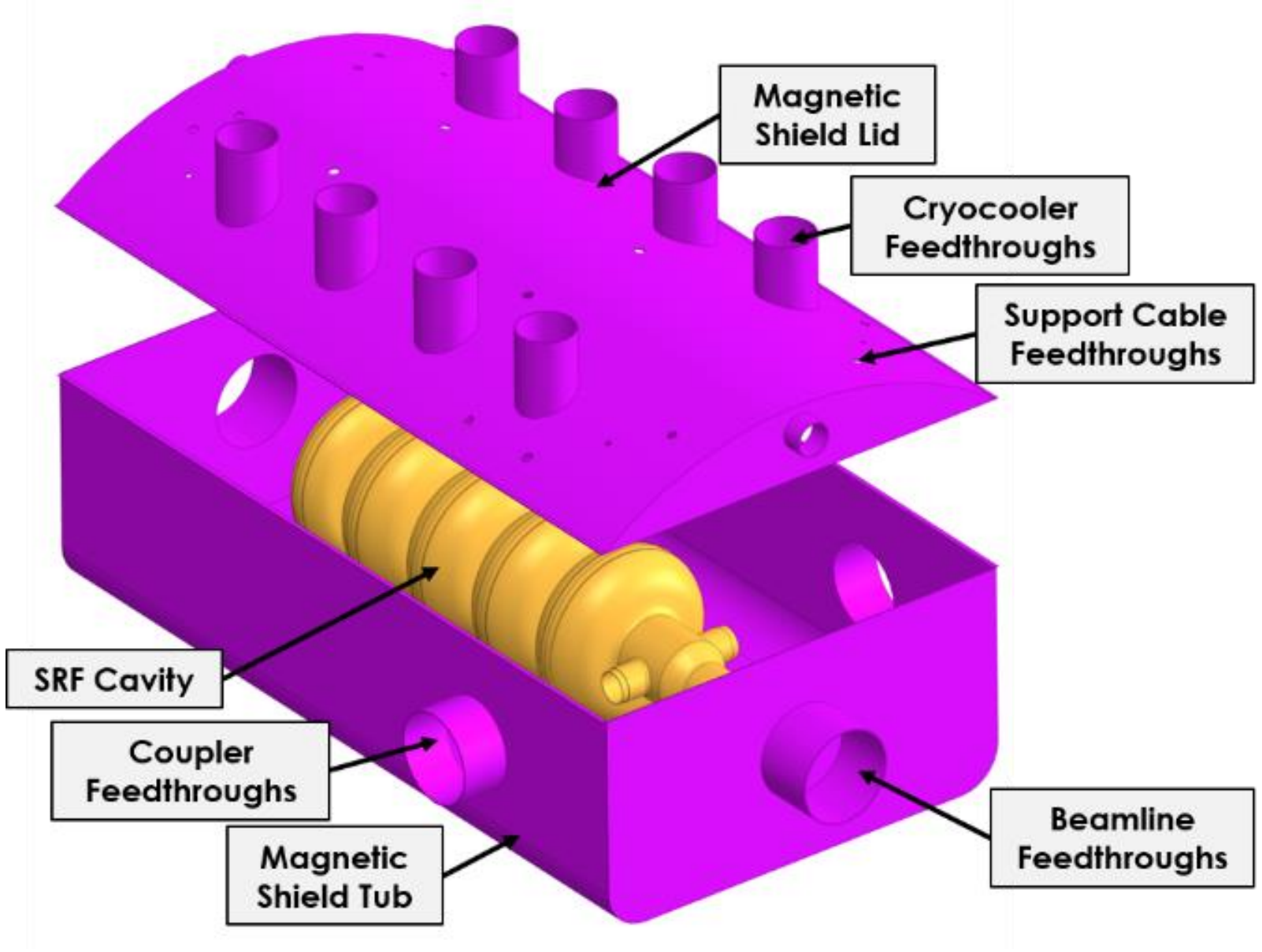

(a)

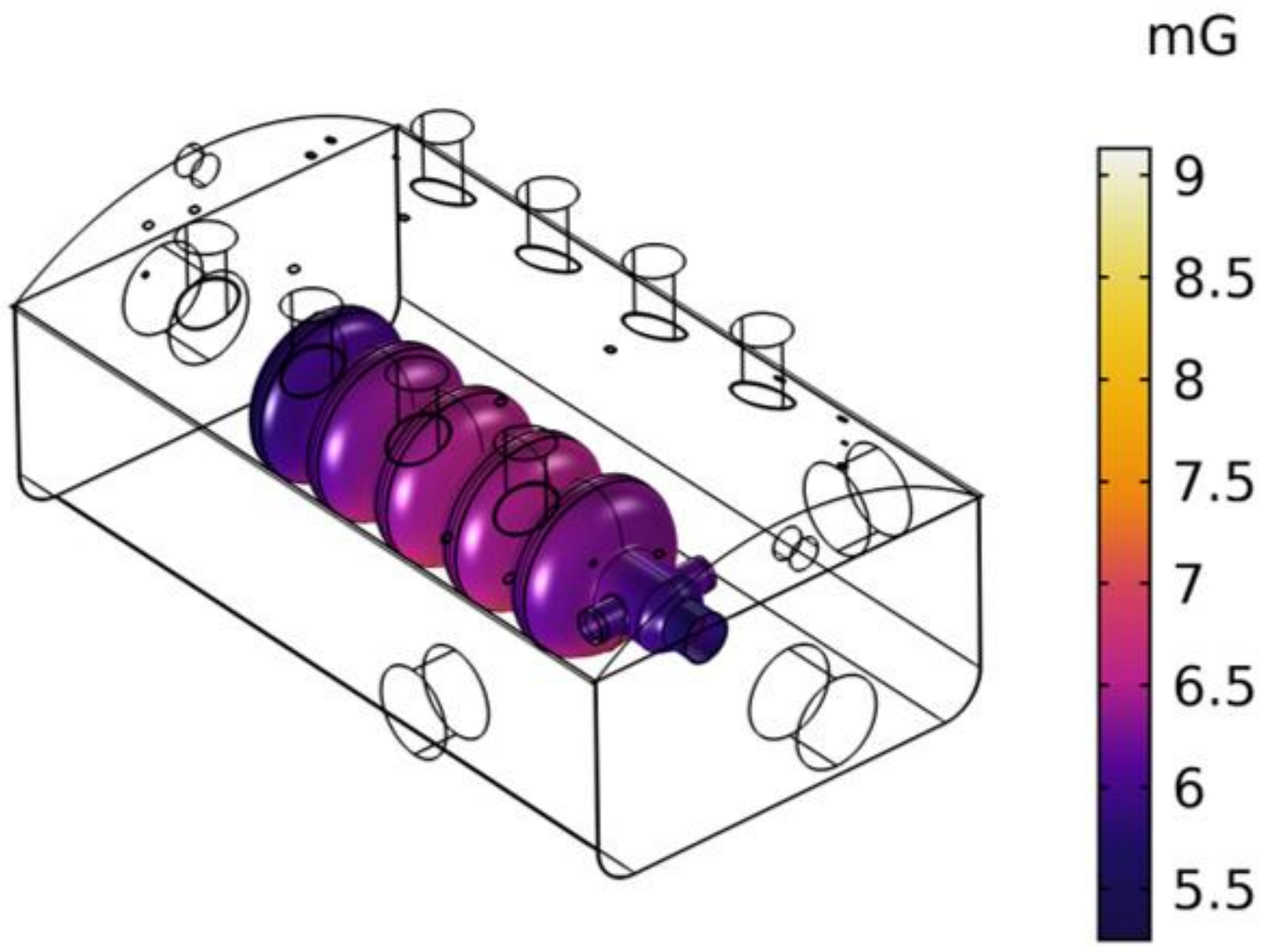

(b)

*Figure 26: (a) The exploded view of the magnetic shield shows that it is composed of two halves, a top lid and a bottom tub which are attached at a small interface flange during assembly. Chimneys are attached around each feedthrough to improve shielding (b) Magnetic flux density displayed along the surface of the SRF cavity, in the case where the beamline is oriented toward the North.*

<u>Vacuum vessel</u>

The cryomodule vacuum vessel shown in Figure 27 is made of 316L stainless steel and consists of two parts: a bottom tub and a top lid. The lid and the tub are detachable. Vacuum seal is established using an o-ring along the periphery, pressed using bolted connections. The vacuum vessel walls are 5/16” thick and the structure is reinforced on the outside by 3/8”-to-1/2”-thick stiffeners that prevent buckling under external pressure. The total weight of the cryostat vacuum vessel is approximately 462 kg, *i.e.*, 168 kg for the lid and 293 kg for the tub. Vacuum sealing features, flange bolt holes, and other small mechanical details are omitted for the evaluation of the overall structural integrity. These can be taken into account during the final design for manufacturing.

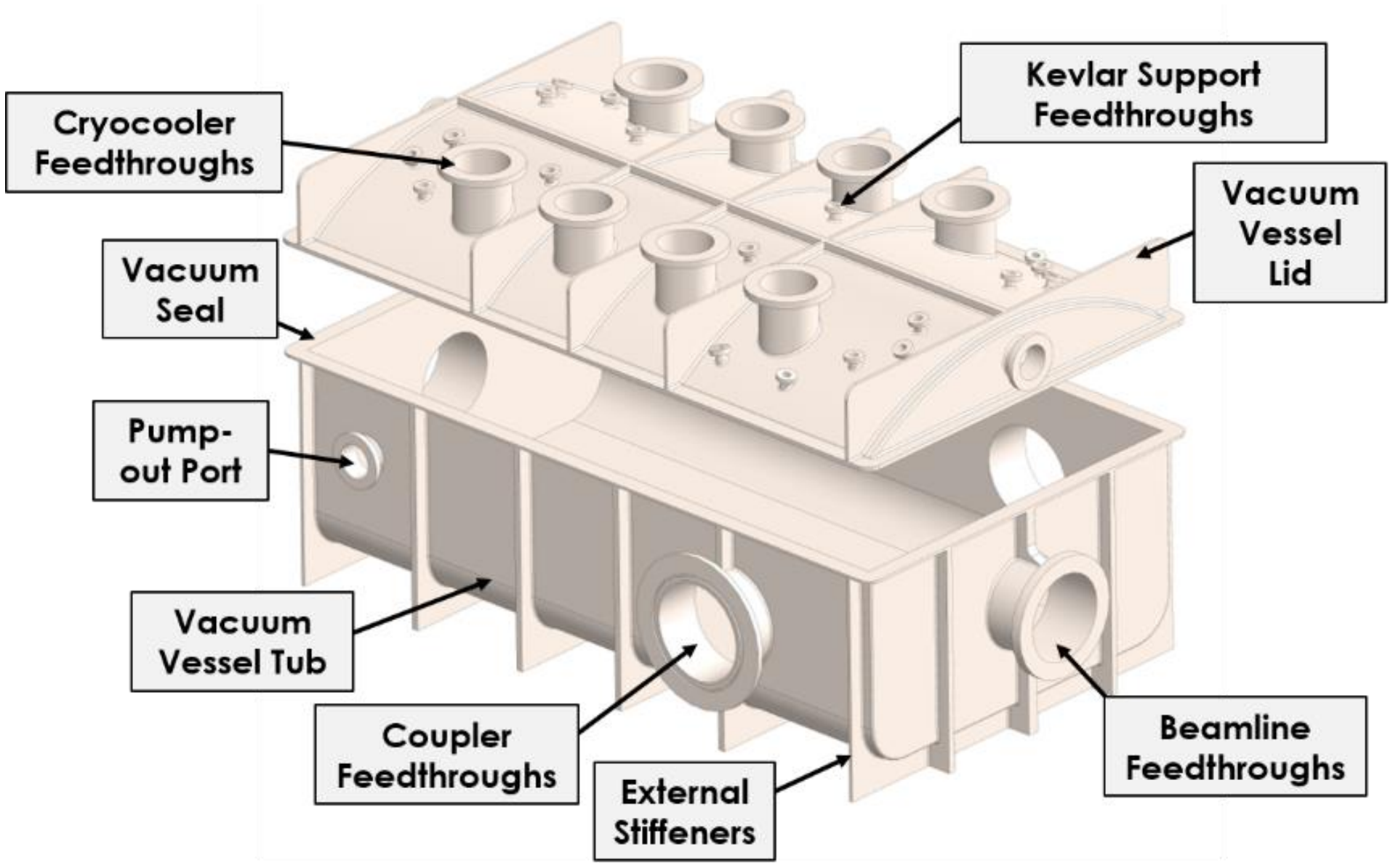

*Figure 27: CAD view of the cryostat vacuum vessel made of a large tub and a lid that are sealed together. External stiffeners are used around the outside of the vessel to reduce stresses and prevent buckling.*

The Solid Mechanics module of COMSOL Multiphysics is used to evaluate stresses and buckling modes of the cryostat. Boundary conditions account for pressure differential across the vacuum vessel wall, gravity forces, the weights of the cavity, cryocoolers, RF couplers, thermal and magnetic shields and beamline components. The material properties are assumed elastic and isotropic.

The approach of the ASME Boiler and Pressure Vessel Code Section VIII Division 2 [31] is used to detrermine the structural adequacy of the vacuum vessel design. The ASME code defines the requirements for protection against three modes of failure, namely protection against plastic collapse, protection against local failure and protection against buckling. To avoid plastic collapse, stresses for 316L Stainless Steel (SA240) should remain below 16.7 ksi, according to Part D, Table 5A of [31]. This criterion is verified using the maximum distortion energy yield criterion, also called von Mises criterion. Von Mises stresses are calculated at all points within the vacuum vessel and are found not to exceed the limit of 16.7 ksi, see Figure 28. Regions of high stresses are located underneath the tub along the middle transversal stiffener and at the contact surfaces between the lid and the tub. The exagerated (130 times) cryostat deformations in Figure 28 indicate that the largest displacements occur at the top and bottom of the vacuum vessel but do not exceed 0.04 inches in amplitude. Furthermore, the risk of local failure is examined in the region of maximum von Mises stress at the bottom of the vacuum vessel. The combined membrane and bending stresses are verified to not exceed a threshold value of 25 ksi, which, according to ASME code, ensures that local failure does not occur.

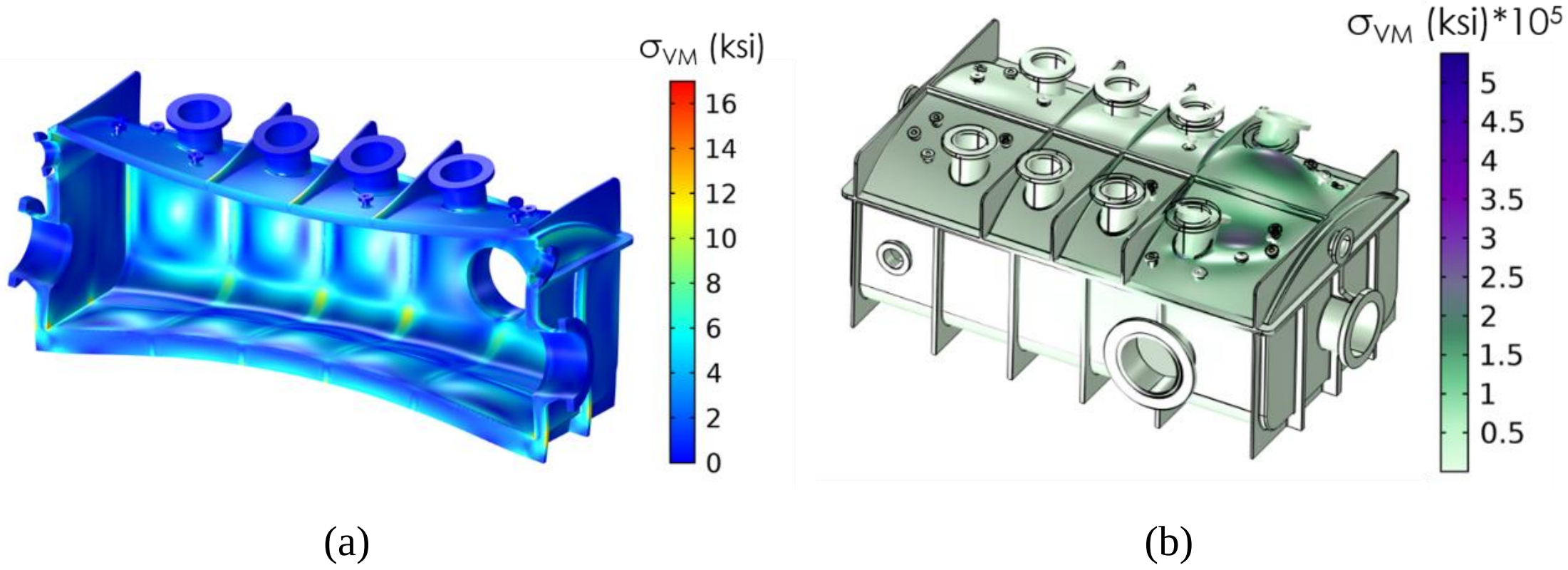

*Figure 28: a) Distribution of von Mises stresses in the vacuum vessel (cross-section view) with exaggerated displacements by a factor 130; b) Von Mises stresses in the vacuum vessel when the first buckling failure mode occurs on one side of the lid. The loading required for this mode of failure is approximately 64 times larger than under normal loading.*

To assess the risk of structural buckling when the vessel is evacuated to a vacuum, a bifurcation buckling analysis is performed using the same boundary conditions as those used for the static stress analysis above. The von Mises stresses when the first buckling mode occurs are showed in Figure 28(b). A design safety factor, $\Phi_B$, is defined as the ratio of local von Mises stress when the vessel collapses by buckling to the local von Mises stress at the applied loading. The minimum allowable design safety factor is determined by $\Phi_B = 2/\beta_{critical}$, where $\beta_{critical} = 0.124$ for external pressure loading. Therefore, it is required that $\Phi_B$ remain greater than 16.1 to avoid buckling failure. The design safety factor is found to exceed 63.4 in the whole vacuum vessel and therefore the vessel will not buckle when evacuated.

In conclusion, the structural analysis demonstrates that the vacuum vessel well exceeds the strength requirements from ASME Section VIII Division 2.

**f. Cryomodule assembly procedure**

The cavity assembly procedure is pictorially represented in Figure 29 and Figure 30. Figure 29(a) shows the cavity assembly, including the thermal links, beamline bellows and valves, and vacuum side of the RF couplers, that is prepared in a clean room prior to integration with the cryomodule. This step ensures no contamination of the beamline. The top assembly of vacuum vessel, thermal shield, and magnetic shield including cryocoolers and beamline suspension components (Kevlar straps) is depicted in Figure 30(b).

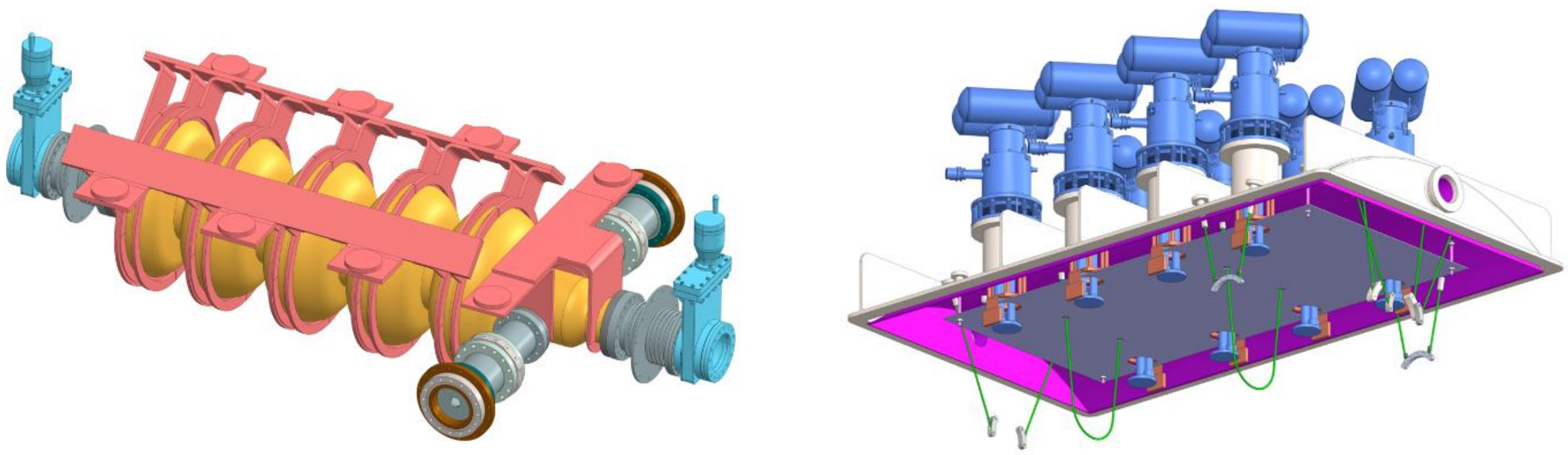

(a) (b)

*Figure 29: CAD views of (a) the beamline assembly prior to installation in the cryocooler and (b) the beamline assembly suspended on the lid of the vacuum vessel.*

The cryomodule assembly sequence is illustrated in Figure 30. All the shields and beamline components are suspended under the lid. To begin, all supporting cables and threaded rods are connected to the lid. The top plates of the magnetic and thermal shields, including the magnetic shield chimneys fitting in the cryocooler ports, are inserted under the lid and attached to the titanium threaded rods. Next, the evacuated cavity, with beamline bellows, gate valves and RF couplers attached, is lifted under the lid and connected to the kevlar straps (Figure 30(a)). The cryocoolers with the cylindrical thermal shield extension attached to the 1$^{st}$ cooling stages are inserted through the vacuum ports and connected to the vacuum vessel using vacuum bellows. The 1$^{st}$ stages of the cryocoolers are connected to the the thermal shields and the 2$^{nd}$ cooling stages to the cavity thermal links. The next step involves assembling the rest of the thermal shield and wrapping the shield in a 30-layer thermal-insulating blanket. The RF couplers are hung from the vacuum vessel top plate using kevlar straps (Figure 30(b)). The side and bottom panels of the magnetic shield are then assembled and attached inside the vacuum vessel tub (Figure 30(c)). Finally, the lid and all hanging components are lowered into the vacuum tub. The horizontal magnetic shield chimneys and the external flanges can be attached to the RF couplers (Figure 30(d)). The cryomodule is designed to be fully disassembled and put back together if needed and has no hermetic welds.

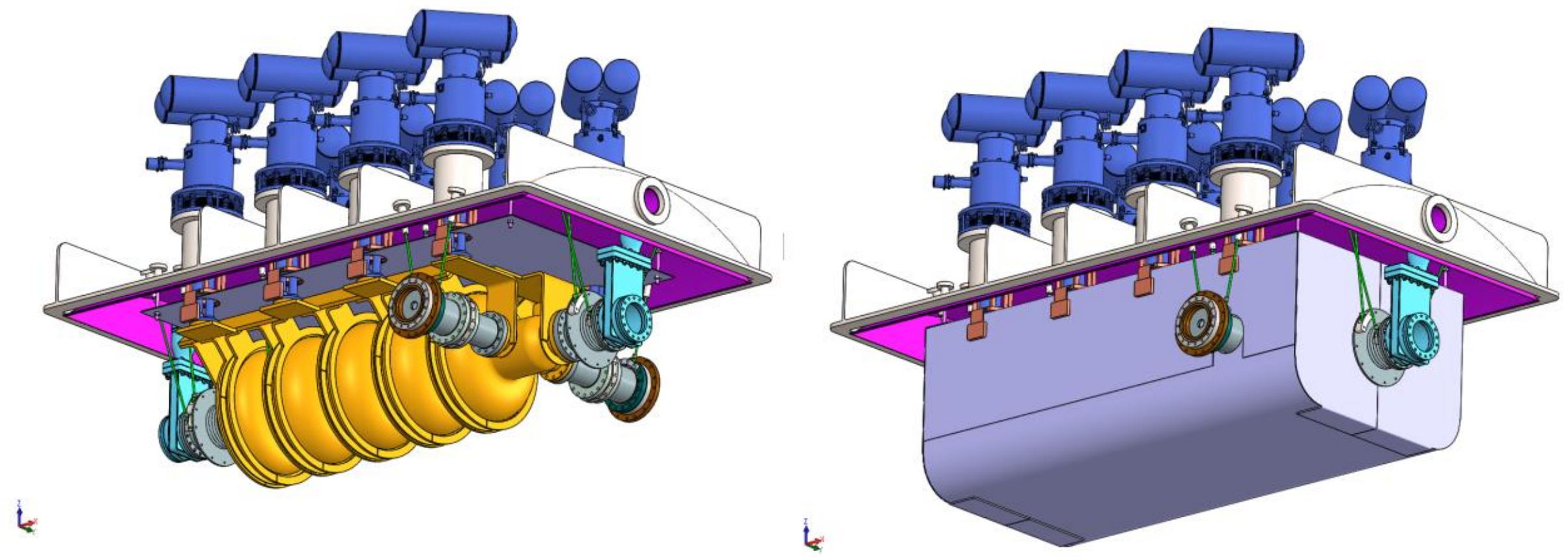

(a) (b)

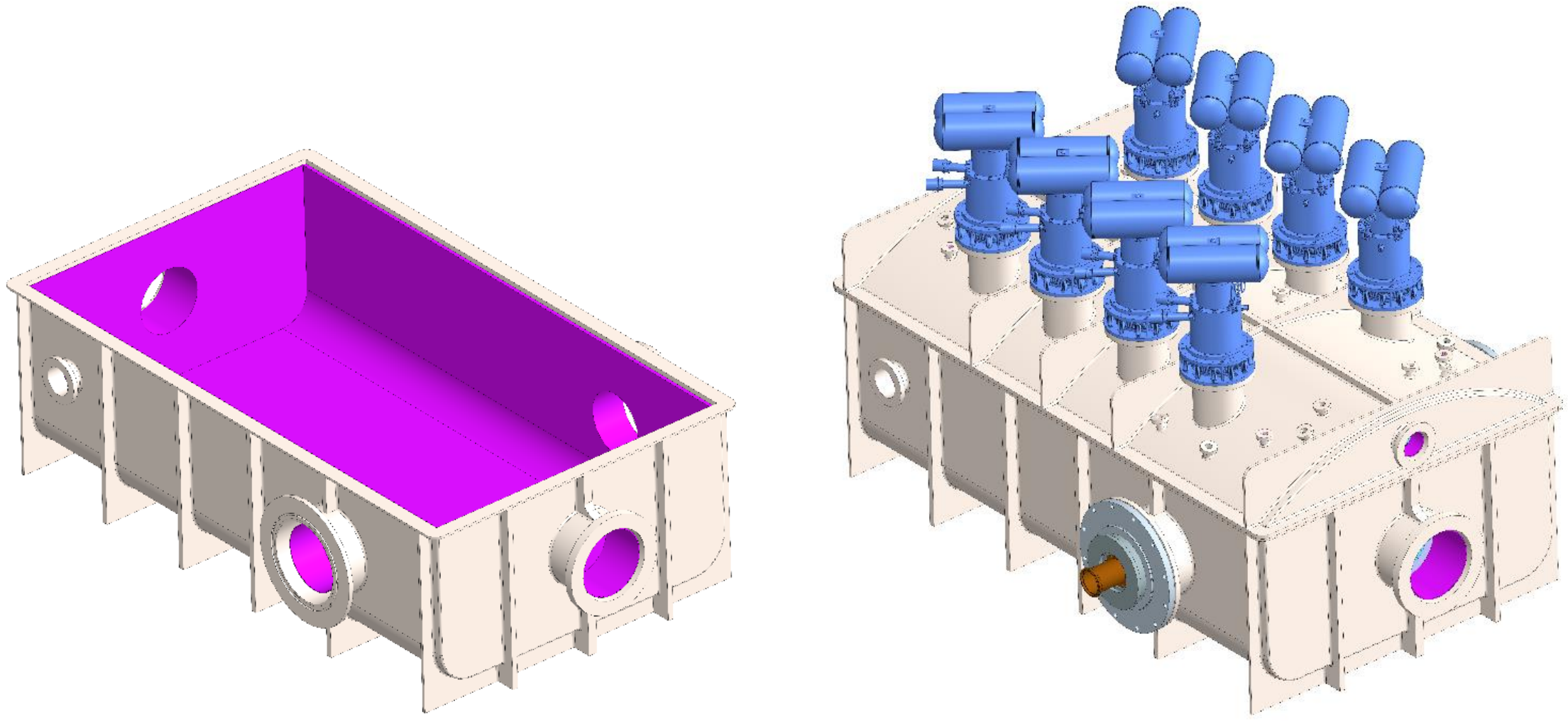

(c) (d)

*Figure 30: Cryomodule assembly sequence - (a) installation of the beamline assembly and cryocoolers on the vacuum vessel lid (b) installation of the the thermal shield (c) Installation of the bottom half of the magnetic shield inside the vacuum vessel tub and (d) fully assembled cryomodule.*

## 3. Accelerator wall-plug efficiency and cost analysis

### a. Wall plug to beam efficiency

The estimated wall-plug-to-beam efficiency of the 1 MW, 10 MeV electron beam is 41% as represented by the power flow diagram in Figure 31. The electric power consumption is dominated by the 1 MW RF source (klystron, for instance) and the associated auxiliary systems (chillers, power supply, RF couplers, etc.), with a combined wall-plug-to-RF-power efficiency of 52%. Losses in the beamline and the beam delivery system are assumed to account for 5% of the input RF power. Additional 20 kW electrical power is required for water cooling for the eight cryocoolers. In total, 2.32 MW of total wall-plug electric power is needed to produce a ~1 MW electron beam. The energy efficiency of the proposed system is comparable to the 1 MW, 1 MeV electron beam accelerator of Ciovati *et al.* [13]. This is because the energy consumption of a MW-class RF linac is dominated by the RF power source. The additional cryocooling capacity required for operation of the present 10-MeV SRF cavity has a marginal impact on the overall power consumption.

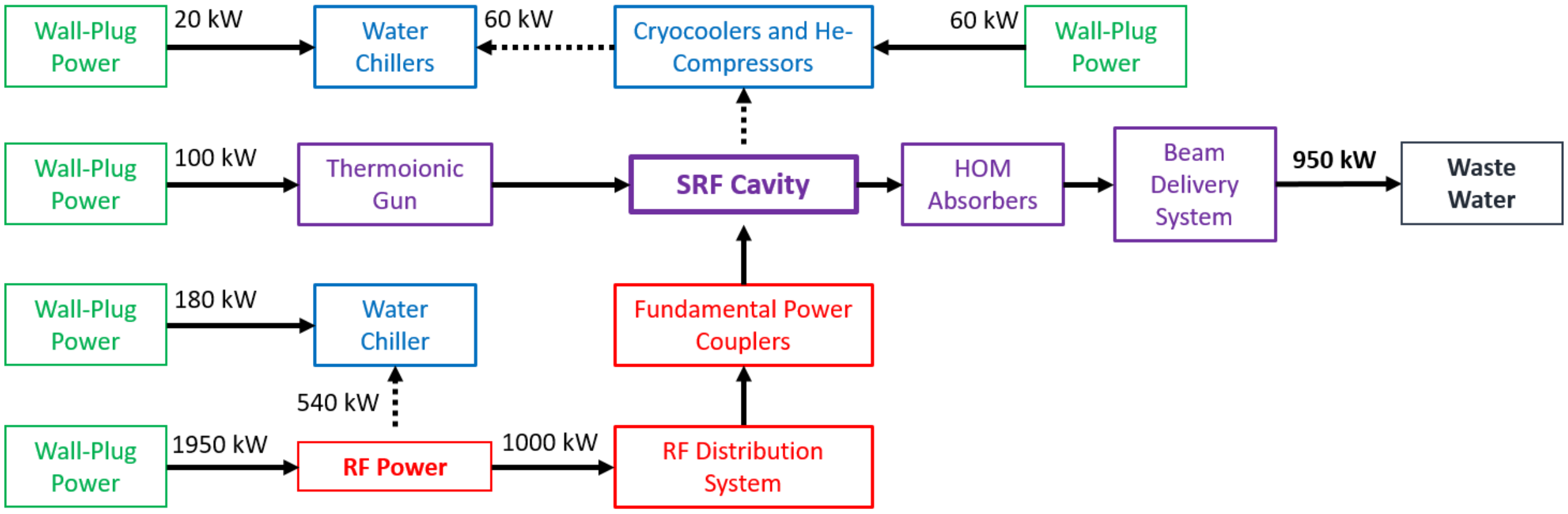

*Figure 31: Power flow diagram for the 10 MeV, 1 MW SRF accelerator.*

### b. Capital expense

The following is a detailed evaluation of the cryomodule capital expenditure (CapEx). The costs of the individual system components are provided by commercial suppliers, manufacturers, and machine shops. The CapEx of the main accelerator cryomodule is projected at approximately $1.55M or $1.62 per watt of beam power. The cost breakdown per system component is presented in Figure 32. The cost of the beamline assembly, composed of the 650 MHz $Nb_3Sn$ cavity, the RF couplers and additional beamline components (valves, bellows and HOM absorber), represents ~50% of the cryomodule CapEx. The eight cryocoolers and the individual helium compressors sum up to approximately $500k, which is 32% of the cryomodule CapEx. The cost of fabrication of the vacuum vessel and the thermal and magnetic shields is $181k, which represents 12% of the cryomodule CapEx. The cost of labor associated with the installation of the beamline and the assembly of the cryomodule is not accounted for in Figure 32.

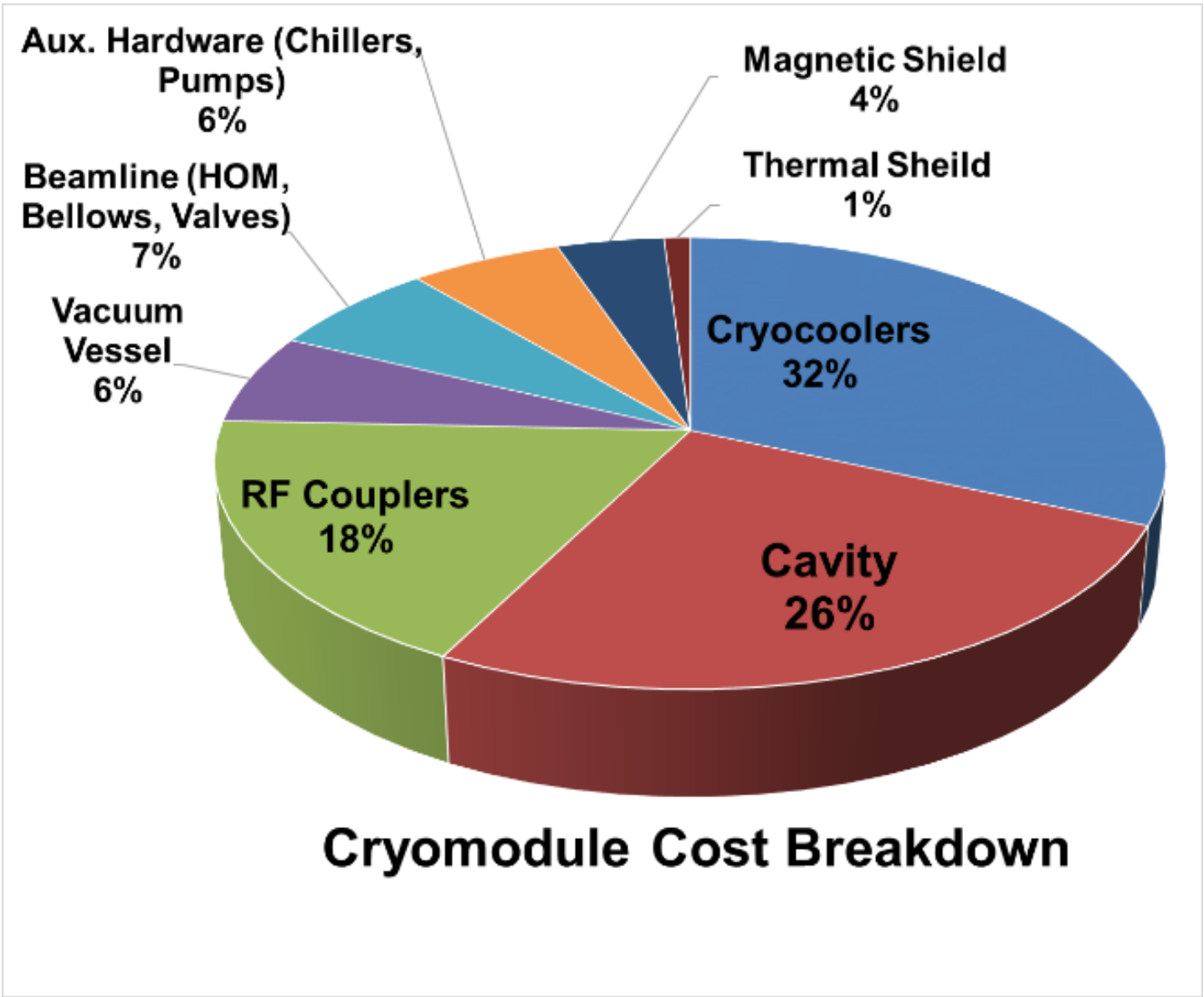

*Figure 32: Cost breakdown of the SRF cryomodule assembly.*

The CapEx of the 1-MW, 1-MeV SRF linac, including the RF source (klystron), the electron gun, the beam delivery system, and the beam diagnostics, was estimated to be $4.5M by Ciovati *et al.* [13]. Based on Ciovati *et al.*'s costing, the CapEx of the present 1 MW, 10 MeV electron-beam accelerator is projected to

be ~$5.1M. Table 7 presents a component wise breakdown of the accelerator CapEx. Ciovati et al.'s 1 MeV linac and the present 10 MeV linac are thought to mostly differ in the design of the cryomodule, the remaining systems being similar or identical. The 10 MeV cryomodule in this study is 67% more expensive than the 1-MeV cryomodule of Ciovati *et al*. The RF power source remains the largest expenditure in the accelerator CapEx. Ciovati *et al.* also evaluated the cost of a 1 MW klystron and all auxiliary systems (power supplies, controllers, RF power distribution, water chillers, etc.) to $3.2M, which exceeds the cost of the 10 MeV cryomodule by a factor two. Alternatives to klystrons such as multi-beam inductive output tubes (MBIOTs) and solid-state RF drivers do not offer any substantial economic or performance benefits at this time. Significant developments in MW-class UHF CW RF source technologies are needed to lower the capital cost and increase the RF efficiency. MW-class RF power sources based on the combined power of multiple low-cost CW magnetrons are in development [32]. This technology has the potential to reduce the cost of industrial MW-class SRF accelerators.

*Table 7: Capital cost (unit: k$) of the 1 MW, 10 MeV SRF accelerator.*

| | |
|---|---|
| ***Total*** | **5,134** |
| ***1 MW RF Power Source [13]*** | **3,200** |
| ***Cryomodule*** | **1,554** |
| *Cryocoolers w/ He Compressors* | 492 |
| *650MHz $Nb_3Sn$ Cavity* | 402 |
| *RF Couplers* | 282 |
| *Vacuum Vessel* | 100 |
| *Beamline (HOM, Bellows, Valves)* | 104 |
| *Auxiliary Hardware (Chillers, Pumps)* | 93 |
| *Magnetic Shield* | 65 |
| *Thermal Shield* | 16 |
| ***Electron Injector [13]*** | **217** |
| ***Beam Delivery System [13]*** | **125** |
| ***Beam Diagnostics & Controls [13]*** | **38** |

## c. Infrastructure and accelerator operating expense

The operating expense (OpEx) of the proposed 10 MeV accelerator, detailed in Table 8, is extrapolated from the costing analysis of Ciovati *et al.* [13] and estimated at $278 per hour of operation. The cost of infrastructure installation (radiation shielding, material delivery system, etc.) is accounted separately from the accelerator CapEx. Ciovati *et al.* estimated the total cost of infrastructure to $2.75M. The present 10 MeV electron beam accelerator necessitates additional shielding than at 1 MeV. The absorbed doses in the forward direction for 1 MeV and 10 MeV electron beam are approximately $2.75 \times 10^{-3}$ rads.m$^2$/hr.kW and $3.00 \times 10^{-5}$ rads.m$^2$/hr.kW, respectively [33]. Therefore, operation of a 10 MeV linac is anticipated to require approximately 20% more shielding (e.g., concrete wall thickness) than for a 1 MeV accelerator of equivalent beam power. As a result, the cost of infrastructure for operation of a 1 MW, 10 MeV linac is projected to be $3.0M. The total cost of fabrication and installation of a 1 MW, 10 MeV SRF linac facility is therefore estimated to be ~$8.1M. Calculations with a 20% loan investment with 15-year amortization are listed in Table 8.

Similar to Ciovati *et al.*, the operating cost is derived under the assumption of high usage of the linac to 8,000 hours per year. The remaining downtime is allocated to maintenance operations. The average annual cost of maintenance is estimated at 2% of the capital cost of the linac facility, which results in an estimated annual maintenance budget of $163k. The cost of electric power consumption is based on an electricity rate of $0.07/kWh, which results in $162 per hour of operation. The linac systems are designed with closed loop water systems and air-cooled chillers. As a result, water consumption is considered negligible. The industrial linac is envisioned as a turn-key system requiring minimal supervision and no specialized or dedicated personnel. Therefore, labor is not accounted for in the operating cost.

### d. Wastewater processing cost

The wastewater processing cost is defined in units of ¢/ton/kGy and represents the cost of a unit dose of 1 kGy deposited in 1 ton of the material. The processing cost of the present 10 MeV linac is estimated at 13.5 ¢/ton/kGy, which is only 6% higher than for the 1 MeV linac of Ciovati *et al.* [13]. This analysis demonstrates that a 10 MeV SRF linac is not excessively more expensive than a 1 MeV system of the same power capacity. For wastewater treatment where dosage of 1-to-4 kGy may be required, the present system could offer a processing capacity of 3-to-12 MGD for a cost of $500-to-$2,000 per mega-gallon of water. For applications requiring significant higher dosage such as 50 kGy for medical waste sterilization, the present linac has the potential to process 48 tons of waste materials per hour at a cost of $5.8 per ton.

*Table 8: Capital investment and estimated cost of operating the 1 MW, 10 MeV SRF accelerator.*

| | |
|---|---|
| ***Capital Investment*** | |
| ***SRF Accelerator*** | $5.13M |
| ***Infrastructure*** | $3.00M |
| ***Investment (20%)*** | $1.63M |
| *Amortization (15 yrs @ 8%)* | $760k |
| ***Operating Cost*** | |
| **Power ($/W)** | $162 /hr |
| **Maintenance** | $163k /yr |
| **Total operating cost** | **$278 /hr** |
| **Processing Cost (¢/ton/kGy)** | **13.5** |

## 4. Summary and outlook

We presented detailed beam dynamics, RF, thermal, and engineering design of a 10 MeV, 1 MW average power e-beam accelerator driven by a room temperature pre-accelerator and a conduction-cooled SRF accelerator cryomodule. The technical design is supplemented by a detailed analysis of capital/construction and operation cost of the e-beam accelerator. The analysis determined that the capital cost is around $8/watt of beam power while ~13.5 ¢/ton/kGy is required for irradiation processing. We consider the 10 MeV, 1 MW accelerator size of 4 m x 2 m x 2 m to be a compact one that can be conveniently set up at municipal and industrial wastewater treatment facilities. While one accelerator unit can treat up to 12 MGD of wastewater, the installation can be easily scaled up for higher volumes by deploying multiple accelerator units.

While the simulation-based design produced in this work appears to be technically feasible as well as cost appealing, a few areas need further practical development. These include production of high $Q_0$ multicell $Nb_3Sn$ cavities, continued research on conduction cooling techniques that better thermalize the cavity with the cryocoolers, probing and suppressing microphonics that can result from cryocooler vibration, and more. As noted in Figure 31, the RF power source is the dominant consumer of electrical power required to drive the e-beam accelerator. Further research and development on RF sources of potentially higher wall-plug to RF efficiency should be undertaken for lowering the overall accelerator operating cost. Since the RF power source is also expected to be the major capital cost driver (Table 7), it is also essential to explore lower-cost alternatives for manufacturing the power sources. Therefore, a low-cost high-efficiency RF power source development is key to fully exploiting SRF accelerator technology for environmental applications.

## Acknowledgement

Research funded by an Accelerator Stewardship award to R.C. Dhuley by US Department of Energy, Office of Science, Office of High Energy Physics.

This manuscript has been authored by Fermi Research Alliance, LLC under Contract No. DE-AC02-07CH11359 with the U.S. Department of Energy, Office of Science, Office of High Energy Physics.